\documentclass{elsart}%
\usepackage{amsmath}
\usepackage{amssymb}
\usepackage{graphicx}
\usepackage{amsfonts}%
\providecommand{\U}[1]{\protect\rule{.1in}{.1in}}
\begin{document}
\begin{frontmatter}
\title{Verification of bootstrap conditions for amplitudes with
quark exchanges in QMRK}
\thanks[RFBR]{Work supported in part by the Russian Fund of Basic Researches,
project 04-02-16685-a, and in part by the Dynasty
Foundation and Russian Science Support Foundation.}
\author{A.V.~Bogdan}
\ead{A.V.Bogdan@inp.nsk.su}
and \author{A.V.~Grabovsky}\ead{grab@nsunet.ru}
\address{ Institute for Nuclear Physics, 630090 Novosibirsk, Russia\\
and Novosibirsk State University, 630090 Novosibirsk, Russia}
\begin{abstract}
The compatibility of the multi--Regge form  of QCD
amplitudes in the quasi--multi--Regge kinematics (QMRK)
and the s--channel unitarity imposes some constraints on
the effective jet--production vertices. We demonstrate
that these constraints known as bootstrap conditions are
satisfied for the amplitudes with the Reggeized quark
exchanges.
\end{abstract}
\end{frontmatter}

\section{Introduction}

This article continues the development of the quark Reggeization theory
\cite{FS} in QCD. A noticeable progress has been recently achieved here, in
particular, the quark Regge trajectory in the next--to--leading approximation
(NLA) in D dimensions was found \cite{BD-DFG,Bogdan_F_1} and the
next--to--leading order (NLO) corrections to the effective vertices appearing
in the leading logarithmic approximation (LLA) were calculated
\cite{FF01,KLPV}. All these results were obtained assuming the reggeized form
for amplitudes in the multi--Regge kinematics (MRK) in the NLA. It is clear
that this assumption, called the quark Reggeization hypothesis, must be
proved. However, in the NLO it is tested only in $\alpha_{s}^{2}$ order
\cite{BD-DFG,Bogdan_F_1} so far. Moreover, its complete proof in the LLA for
any quark--gluon inelastic process in all orders of\emph{ }$\alpha_{s}$ was
given only recently \cite{Bogdan:2006af}. This proof \ is based on the
relations required by compatibility of the multi--Regge form of QCD amplitudes
with the $s$--channel unitarity (bootstrap relations). The fulfillment of the
bootstrap relations is secured by several conditions (bootstrap conditions) on
Reggeon vertices and trajectories. An analogous proof can (and has to) be
constructed in the NLA as well.

The only kinematics essential in the LLA is MRK, which means that all
particles produced in a high--energy process have limited transverse momenta
and are well separated in rapidity space. In the NLA, production amplitudes in
this kinematics can be obtained by taking one of the effective vertices or
Regge trajectory in the NLO. But in the NLA another, quasi--multi--Regge
kinematics (QMRK) becomes also important. In this case one of the produced
particles is replaced by a jet containing two particles with similar
rapidities. At this moment all multi--particle Reggeon vertices required in
the NLA are obtained \cite{Lipatov:2000se}. Therefore, a proof of the quark
Reggeization hypothesis concerning the QMRK may be given.

In this paper we prove the quark Reggeization in the QMRK. Our method is the
direct continuation of the one used to prove this hypothesis in the LLA. The
bootstrap conditions for amplitudes in the QMRK are the same as in the LLA
with the substitution of jet production vertices for particle production ones.
Hereafter we demonstrate that all these conditions are fulfilled. For
convenience we work in the operator formalism, which was introduced in
\cite{FFKP} and extended to inelastic amplitudes and quark exchanges in
\cite{Bogdan:2006af}.

The paper is organized as follows. The next section contains all necessary
denotations and the definition of the QMRK. Section 3 presents the bootstrap
conditions in operator formalism. In section 4 we prove these conditions for
impact factors and Reggeon--Reggeon--jet (RRJ) effective vertices. Section 5
concludes the paper.

\section{Quasi--multi--Regge form of QCD amplitudes}

Considering the QMRK we talk about a multiparticle production amplitude as
about the amplitude of jet production where one of the jets consists of two
particles. Such a jet can be produced either in the fragmentation regions of
initial particles, or in the central region, i.e. with the rapidity far away
from the rapidities of colliding particles. Let us consider the process
$A+B\rightarrow A^{\prime}+P_{1}+....+P_{n}+B^{\prime}$ in the QMRK. Using the
same denotations as in \cite{Bogdan:2006af} we introduce light--cone momenta
$n_{1}$ and $n_{2}$: $\;\;n_{1}^{2}=n_{2}^{2}=0,$ $(n_{1}n_{2})=1$ and denote
$(pn_{2})\equiv p^{+},\;\;(pn_{1})\equiv p^{-}$, so that $pq=p^{+}q^{-}%
+p^{-}q^{+}+p_{\bot}q_{\bot}$. Here the sign $\bot$ means transverse to the
$(n_{1},n_{2})$ plane components. We assume that initial momenta $p_{A}$ and
$p_{B}$ have predominant components along $n_{1}$ and $n_{2}$ respectively.
For generality we do not demand that transverse components $p_{A\bot}$ and
$p_{B\bot}$ should be zero, but assume $|p_{A\bot}^{2}|\sim|p_{B\bot}^{2}|\sim
p_{A}^{2}\sim p_{B}^{2}\ll p_{A}^{+}p_{B}^{-}\;$ and remain limited (not grow)
at $p_{A}^{+}p_{B}^{-}\rightarrow\infty$. For the final jet momenta
$p_{i},\;\;i=0,....,n+1$, we assume the QMRK conditions:
\begin{align}
p_{0}^{-} &  \ll p_{1}^{-}\ll\ldots\ll p_{n}^{-}\ll p_{n+1}^{-}\,,\nonumber\\
p_{n+1}^{+} &  \ll p_{n}^{+}\ll\ldots\ll p_{1}^{+}\ll p_{0}^{+}%
\,,\label{alph-beta}%
\end{align}
where $p_{i\bot}$ are limited. It ensures that the squared invariant masses
$s_{ij}=(p_{i}+p_{j})^{2}$ are large compared with the squared transverse
momenta and invariant masses of the jets. At $i<j$ they have the form
\begin{equation}
s_{ij}\approx2p_{i}^{+}p_{j}^{-}=\frac{p_{i}^{+}}{p_{j}^{+}}(p_{j}%
^{2}-p_{j\bot}^{2})=\frac{p_{j}^{-}}{p_{i}^{-}}(p_{i}^{2}-p_{i\bot}%
^{2})\,,\label{s-ij}%
\end{equation}
and at $i<l<j$ submit to relations
\begin{equation}
s_{il}s_{lj}\approx s_{ij}(p_{l}^{2}-p_{l\bot}^{2})\,.\label{s-ilj}%
\end{equation}
For the momentum transfers $q_{i}$, $i=1,\ldots,n+1$,
\begin{equation}
q_{1}=p_{0}-p_{A}\,,\;\;q_{j+1}=q_{j}+p_{j}\,,\quad(j=1,\ldots,n)\,,\
\end{equation}
we have
\begin{equation}
q_{i}^{2}\approx q_{\bot}^{2}\,.\label{trans}%
\end{equation}
In the LO the amplitude $A_{2\rightarrow n+2}$ of the process $A+B\rightarrow
A^{\prime}+P_{1}+....+P_{n}+B^{\prime}$ has the multi--Regge form
\begin{equation}
A_{2\rightarrow n+2}^{\mathcal{R}}=\bar{\Gamma}_{A^{\prime}A}^{\mathcal{R}%
_{1}}\frac{s_{1}^{\omega_{1}}}{d_{1}}\gamma_{\mathcal{R}_{1}\mathcal{R}_{2}%
}^{P_{1}}\frac{s_{2}^{\omega_{2}}}{d_{2}}.....\gamma_{\mathcal{R}%
_{n}\mathcal{R}_{n+1}}^{P_{n}}\frac{s_{n+1}^{\omega_{n+1}}}{d_{n+1}}%
\Gamma_{B^{\prime}B}^{\mathcal{R}_{n+1}}\,,\label{inelastic quark}%
\end{equation}
where $\bar{\Gamma}_{A^{\prime}A}^{\mathcal{R}}$ and $\Gamma_{B^{\prime}%
B}^{\mathcal{R}}$ are the particle--particle--Reggeon (PPR) effective
vertices, describing particle--particle $P\rightarrow P^{\prime}$ transitions
due to interaction with Reggeons $\mathcal{R}$.$\;$For gluon quantum numbers
in $q_{i}$ channel, $\omega_{i}=\omega_{\mathcal{G}}(q_{i})$ is the gluon
Regge trajectory and $d_{i}\equiv d_{i}(q_{i})=q_{i\bot}^{2}$; for quark
numbers, $\omega_{i}=\omega_{\mathcal{Q}}(q_{i})$ is the quark Regge
trajectory and $d_{i}\equiv d_{i}(q_{i})=m-\hat{q}_{i\bot}.\;$ $\gamma
_{\mathcal{R}_{i}\mathcal{R}_{i+1}}^{P_{i}}$ are the
Reggeon--Reggeon--particle (RRP) effective vertices describing production of
particle $P_{i}$ at Reggeon transitions $\mathcal{R}_{i+1}\rightarrow
\mathcal{R}_{i}$. In order to be definite we do not consider here antiquark
quantum numbers in any of $q_{i}$ channels. It determines the order of the
multipliers in \eqref{inelastic quark}. Nonetheless, our consideration does
not lose generality because amplitudes with quark and antiquark exchanges are
related by charge conjugation.

Since we come to the QMRK replacing one of the particles $P_{i}$ in the MRK
with a pair with fixed invariant mass, QMRK amplitudes have the same form
\eqref{inelastic quark} as LO MRK ones, where one of the vertices
$\gamma_{\mathcal{RR}}^{P}$ or $\Gamma_{P^{\prime}P}^{\mathcal{R}}$ is
substituted with the jet production vertex $\gamma_{\mathcal{RR}}%
^{\{P_{1}P_{2}\}}$ or $\Gamma_{\{P_{1}^{\prime}P_{2}^{\prime}\}P}%
^{\mathcal{R}}$ respectively. Note, that because the QMRK leads to the loss of
a large logarithm in the unitarity relations, energy scales in
(\ref{inelastic quark}) are unimportant in the NLA. Moreover, we need
trajectories and vertices only in the LO there. Assuming similar ordering of
longitudinal components one can obtain the more general multi--jet amplitudes
$A_{2+n_{1}\rightarrow2+n_{2}}^{\mathcal{R}}$ from $A_{2\rightarrow
n+2}^{\mathcal{R}}$ by usual crossing rules. Note, that as in
\eqref{inelastic quark} we can neglect imaginary parts of these amplitudes
since in the QMRK they are next--to--next--to-leading. Therefore, as well as
for the amplitudes in the LO, crossing rules connecting the QMRK amplitudes do
not affect the Regge factors $s_{i}^{\omega_{i}}$.

Hereafter we work in the physical light--cone gauge
\begin{equation}
(e\,p)=(e\,n_{1})=0,\;\;e=e_{\bot}-\frac{(ep)_{\bot}}{p^{-}}n_{1}%
,\label{n1-gauge}%
\end{equation}
where $e$ is the polarization vector of a gluon with momentum $p$.

We use the PPR vertices in the LO in this gauge from \cite{Bogdan:2006af}:
\begin{gather}
\Gamma_{G^{\prime}G}^{\mathcal{G}}=-2g\,p_{G}^{-}{T}_{G^{\prime}%
G}^{\mathcal{G}}(e_{G_{\bot}^{\prime}}^{\ast}e_{G_{\bot}})\,,\quad
\Gamma_{Q^{\prime}Q}^{\mathcal{G}}=g\,\bar{u}_{Q^{\prime}}t^{\mathcal{G}%
}{\gamma^{-}}u_{Q}\,,\nonumber\\
\bar{\Gamma}_{G^{\prime}G}^{\mathcal{G}}=-2g\,p_{G}^{+}{T}_{G^{\prime}%
G}^{\mathcal{G}}(e_{G_{\bot}^{\prime}}^{\ast}e_{G_{\bot}})\,,\quad\bar{\Gamma
}_{Q^{\prime}Q}^{\mathcal{G}}=g\,\bar{u}_{Q^{\prime}}t^{\mathcal{G}}%
{\gamma^{+}}u_{Q}\,,\label{PPRg vertices}\\
\Gamma_{\bar{Q}^{\prime}\bar{Q}}^{\mathcal{G}}=-g\,\bar{\upsilon}_{\bar{Q}%
}t^{\mathcal{G}}{\gamma^{-}}\upsilon_{\bar{Q}^{\prime}}\,,\quad\bar{\Gamma
}_{\bar{Q}^{\prime}\bar{Q}}^{\mathcal{G}}=-g\,\bar{\upsilon}_{\bar{Q}%
}t^{\mathcal{G}}{\gamma^{+}}\upsilon_{\bar{Q}^{\prime}}\,,\nonumber
\end{gather}%
\begin{gather}
\Gamma_{G^{\prime}Q}^{\mathcal{Q}}=-gt^{G^{\prime}}\hat{e}_{G^{\prime}\bot
}^{\ast}u_{Q}\,,\quad\Gamma_{\bar{Q}^{\prime}G}^{\mathcal{Q}}=-gt^{G}\hat
{e}_{G\bot}\upsilon_{\bar{Q}^{\prime}}\,,\nonumber\\
\bar{\Gamma}_{G^{\prime}\bar{Q}}^{\mathcal{Q}}=-g\bar{\upsilon}_{\bar{Q}%
}t^{G^{\prime}}\hat{e}_{{G^{\prime}}\bot}^{\ast}\,,\quad\bar{\Gamma
}_{Q^{\prime}G}^{\mathcal{Q}}=-g\bar{u}_{Q^{\prime}}t^{G}\hat{e}_{G\bot
}\,.\label{PPRq vertices}%
\end{gather}
Here we denote particles and Reggeons by symbols which accumulate all their
quantum numbers. We use the letter $P$ for particles (jets) and the letter
$\mathcal{R}$ for Reggeons independently of their nature, letters $G$ and $Q$
for ordinary gluons and quarks and $\mathcal{G}$ and $\mathcal{Q}$ for the
Reggeized ones. In the gauge (\ref{n1-gauge}) the RRP vertices for gluon,
quark and antiquark production with momentum $p=q_{2}-q_{1}$ at the transition
of Reggeon $\mathcal{R}_{2}$ (with momentum $q_{2}$) to Reggeon $\mathcal{R}%
_{1}$(with momentum $q_{1}$) are as follows \cite{Bogdan:2006af}:
\begin{gather}
\gamma_{\mathcal{G}_{1}\mathcal{G}_{2}}^{G}=2g{T}_{\mathcal{G}_{1}%
\mathcal{G}_{2}}^{G}\,{e_{\perp}^{\ast}}\left(  q_{2\bot}-p_{G\bot}%
\frac{q_{2\bot}^{2}}{p_{G\bot}^{2}}\right)  ,\\
\gamma_{\mathcal{Q}_{1}\mathcal{Q}_{2}}^{G}=-gt^{G}\,{e_{\perp}^{\ast}}\left(
\gamma_{\bot}-2(\hat{q}_{2\bot}-m)\frac{p_{G\bot}}{p_{G\bot}^{2}}\right)
\,,\label{gammaG-n1}%
\end{gather}%
\begin{equation}
\gamma_{\mathcal{G}_{1}\mathcal{Q}_{2}}^{Q}=g\,\,\bar{u}_{Q}\frac{\hat
{q}_{1\bot}}{p_{Q}^{+}}t^{\mathcal{G}_{1}},\qquad\gamma_{\mathcal{Q}%
_{1}\mathcal{G}_{2}}^{\bar{Q}}=-gt^{\mathcal{G}_{2}}\,\frac{\hat{q}_{2\bot}%
}{p_{\bar{Q}}^{-}}\upsilon_{\bar{Q}}\,.
\end{equation}
The particle--jet--Reggeon (PJR) and Reggeon--Reggeon--jet (RRJ) effective
vertices taken from \cite{Fadin:2003av} and \cite{Lipatov:2000se} can be
presented in different forms and our goal is to find the presentation in which
subsequent calculations become trivial. Taking this in mind we introduce a set
of functions $F_{i}$, $K_{i}$, $V_{i}$ (see below eqs.(\ref{func1}%
--\ref{func2}) which determine all the PJR and RRJ effective\ vertices. The
selection of these functions is a nontrivial task and a result of the analysis
of cancellations during the verification of all QMRK bootstrap conditions.

Firstly, we introduce $x_{i}=k_{i}^{-}/k^{-}$, with $k_{i}$ being the momentum
of the final particle and $\ k=k_{1}+k_{2}$ the momentum of the jet, so
$x_{1}+x_{2}\simeq1$, we also use $v=x_{2}k_{1\bot}-x_{1}k_{2\bot}$. The
commonly arising denominators may be expressed via
\begin{equation}
D(p,q)=x_{1}p_{\bot}^{2}+x_{2}q_{\bot}^{2}\quad\text{and }\quad d(p,q)=(x_{1}%
p_{\bot}-x_{2}q_{\bot})^{2}\,.
\end{equation}
We rewrite all RRP and PJR effective vertices in terms of the functions
$G_{i}$, $K_{i}$, $V_{i}$ \ by means of the following identities%
\begin{align}
&  \frac{1}{k_{2\bot}^{2}(k_{1\bot}^{2}-m^{2})}=\frac{x_{2}}{k_{2\bot}%
^{2}(D(k_{2},k_{1})-x_{2}m^{2})}\nonumber\\
&  +\frac{x_{1}}{(k_{1\bot}^{2}-m^{2})(D(k_{2},k_{1})-x_{2}m^{2}%
)}\,,\allowbreak\label{id1}%
\end{align}%
\begin{align}
&  \frac{x_{1}x_{2}}{(d(k_{2},k_{1})-x_{2}^{2}m^{2})(D(k_{2},k_{1})-x_{2}%
m^{2})}\nonumber\\
&  =\frac{1}{(k_{1}+k_{2})_{\bot}^{2}-m^{2}}\left(  \frac{1}{d(k_{2}%
,k_{1})-x_{2}^{2}m^{2}}-\frac{1}{D(k_{2},k_{1})-x_{2}m^{2}}\right)  \,,
\end{align}%
\begin{align}
&  x_{1}^{2}k_{2\bot}^{2}-x_{2}^{2}k_{1\bot}^{2}=(x_{1}-x_{2})d(k_{2}%
,k_{1})-2x_{1}x_{2}(k_{1}+k_{2},v)_{\bot}\nonumber\\
&  =x_{1}x_{2}(k_{2\bot}^{2}-k_{1\bot}^{2})+(x_{1}-x_{2})D(k_{2},k_{1})\,,
\end{align}%
\begin{align}
&  (e_{1},v)_{\bot}(e_{2}k_{2})_{\bot}+(e_{2},v)_{\bot}(e_{1}k_{1})_{\bot
}\nonumber\\
&  =x_{2}(e_{2},k_{1}+k_{2})_{\bot}(e_{1}k_{1})_{\bot}-x_{1}(e_{1},k_{1}%
+k_{2})_{\bot}(e_{2}k_{2})_{\bot}\nonumber\\
&  =x_{1}(e_{2},v)_{\bot}(e_{1},k_{1}+k_{2})_{\bot}+x_{2}(e_{1},v)_{\bot
}(e_{2},k_{1}+k_{2})_{\bot},
\end{align}%
\begin{equation}
\bar{u}_{k_{1}}\hat{n}_{1}\left(  \frac{\hat{k}_{1\bot}+m}{x_{1}}\gamma_{\bot
}^{\mu}+\gamma_{\bot}^{\mu}\frac{\hat{k}_{2\bot}+m}{x_{2}}\right)
\upsilon_{k_{2}}=2k^{-}\,\bar{u}_{k_{1}}\gamma_{\bot}^{\mu}\upsilon_{k_{2}}\,,
\end{equation}%
\begin{equation}
\bar{u}_{k_{1}}\hat{n}_{1}\left(  \frac{\hat{k}_{1\bot}+m}{x_{1}}\gamma_{\bot
}^{\mu}+\gamma_{\bot}^{\mu}\frac{\hat{k}_{2\bot}-m}{x_{2}}\right)  u_{k_{2}%
}=2k^{-}\,\bar{u}_{k_{1}}\gamma_{\bot}^{\mu}u_{k_{2}}\,,\label{id2}%
\end{equation}
where $e_{1}$and $e_{2}$ are the polarization vectors of emitted gluons. Our
functions are:
\begin{align}
F_{1}^{\mu}(k_{2},k_{1}) &  =\hat{e}_{1\bot}\left(  2x_{1}k_{2\bot}^{\mu
}-x_{2}(\hat{k}_{1\bot}+m)\gamma^{\mu}\right)  \,,\nonumber\\
\tilde{F}_{1}^{\mu}(k_{2},k_{1}) &  =\left.  F_{1}^{\mu}(k_{2},k_{1}%
)\right\vert _{1\leftrightarrow2}\,,\label{func1}%
\end{align}%
\begin{equation}
F_{2}^{\mu}(k_{2},k_{1})=\gamma^{\mu}\left(  2x_{1}(e_{2},v)_{\bot}%
+x_{2}\,\hat{e}_{2\bot}\,(\hat{v}+x_{2}m)\right)  ,
\end{equation}%
\begin{align}
F_{3}^{\mu}(k_{2},k_{1}) &  =\hat{e}_{1\bot}\left(  2v^{\mu}-x_{2}%
\,\gamma_{\bot}^{\mu}\,(\hat{v}-x_{2}m)\right)  \,,\\
\tilde{F}_{3}^{\mu}(k_{2},k_{1}) &  =\hat{e}_{2\bot}\left(  -2v^{\mu}%
+x_{1}\,\gamma_{\bot}^{\mu}\,(\hat{v}+x_{1}m)\right)  \,,
\end{align}%
\begin{align}
F_{4}^{\mu}(k_{2},k_{1}) &  =\hat{e}_{2}\left(  \gamma^{\mu}\,(\hat
{v}+m)-2x_{1}v{}^{\mu}\right)  \,,\\
\tilde{F}_{4}^{\mu}(k_{2},k_{1}) &  =\hat{e}_{1}\left(  -\gamma^{\mu}%
\,(\hat{v}-m)+2x_{2}v^{\mu}\right)  \,,
\end{align}%
\begin{equation}
F_{5}(k_{2},k_{1})=\,(\hat{k}_{1\bot}+m)(\hat{k}_{2\bot}+m)\,,
\end{equation}%
\begin{align}
K_{1}^{\mu}(k_{2},k_{1}) &  =\frac{2x_{1}\hat{e}_{1\bot}(\hat{k}_{1\bot
}+m)k_{2\bot}^{\mu}}{(D(k_{2},k_{1})-x_{2}m^{2})(k_{1\bot}^{2}-m^{2})}\,,\,\\
\,\tilde{K}_{1}^{\mu}(k_{2},k_{1}) &  =\left.  K_{1}^{\mu}(k_{2}%
,k_{1})\right\vert _{1\leftrightarrow2}%
\end{align}%
\begin{equation}
K_{2}^{\mu}(k_{2},k_{1})=\frac{2x_{2}\hat{e}_{1\bot}(\hat{k}_{1\bot
}+m)k_{2\bot}^{\mu}}{(D(k_{2},k_{1})-x_{2}m^{2})k_{2\bot}^{2}}\,,\quad
\tilde{K}_{2}^{\mu}(k_{2},k_{1})=\left.  K_{2}^{\mu}(k_{2},k_{1})\right\vert
_{1\leftrightarrow2}\,,
\end{equation}%
\begin{equation}
K_{3}^{\mu}(k_{2},k_{1})=\frac{4x_{1}k_{1\bot}^{\mu}(e_{2},k_{2})_{\bot}%
}{D(k_{2},k_{1})k_{1\bot}^{2}}\,,\quad\tilde{K}_{3}^{\mu}(k_{2},k_{1}%
)=\frac{4x_{2}\,k_{1\bot}^{\mu}(e_{2},k_{2})_{\bot}}{D(k_{2},k_{1})k_{2\bot
}^{2}}\,,
\end{equation}%
\begin{equation}
V_{1}^{\mu}(k_{2},k_{1})=2x_{1}x_{2}\left(  e_{1}e_{2}\right)  _{\bot}v^{\mu
}-2x_{1}(e_{2},v)_{\bot}e_{1\bot}^{\mu}-2x_{2}(e_{1},v)_{\bot}e_{2\bot}^{\mu
}\,,
\end{equation}%
\begin{align}
&  e_{1\bot}^{\mu}V_{2}^{\mu}(k_{2},k_{1})=x_{1}x_{2}\left(  e_{1}%
e_{2}\right)  _{\bot}(k_{2\bot}^{2}-k_{1\bot}^{2})\nonumber\\
&  +2\,(e_{1},v)_{\bot}(e_{2}k_{2})_{\bot}+2(e_{2},v)_{\bot}(e_{1}%
,k_{1})_{\bot}\,,
\end{align}%
\begin{equation}
V_{3}^{\mu}(k_{2},k_{1})=2x_{1}x_{2}\hat{e}_{2\bot}v^{\mu}-2x_{1}%
(e_{2},v)_{\bot}\gamma_{\bot}^{\mu}-2x_{2}\hat{v}_{\bot}e_{2\bot}^{\mu
}\,,\label{func2}%
\end{equation}
where $1\leftrightarrow2$ means $k_{1}\leftrightarrow k_{2}$, $x_{1}%
\leftrightarrow x_{2}$, $e_{1}\leftrightarrow e_{2}$.\newline Notice that
\begin{equation}
e_{1\bot}^{\mu}\tilde{F}_{3,4}^{\mu}(k_{2},k_{1})=\left.  e_{2\bot}^{\mu
}F_{3,4}^{\mu}(k_{2},k_{1})\right\vert _{1\leftrightarrow2}%
\end{equation}
and
\begin{equation}
\left.  e_{1}^{\mu}\tilde{K}_{3}^{\mu}(k_{2},k_{1})\right\vert
_{1\leftrightarrow2}=e_{1}^{\mu}K_{3}^{\mu}(k_{2},k_{1})\,.
\end{equation}
The functions $V_{1}\left(  k_{2},k_{1}\right)  ,\,e_{1\bot}^{\mu}V_{2}^{\mu
}\left(  k_{2},k_{1}\right)  $ are antisymmetric under $(1\leftrightarrow2)$
replacement. One can see that $V_{3}^{\mu}(k_{2},k_{1})$ equals $V_{1}^{\mu
}(k_{2},k_{1})$, where $e_{1\bot}$is changed to $\gamma_{\bot}$. We also need
some of these functions with the substitution $e_{1}\rightarrow n_{1}$. We
mark them with a "$-$" sign, e.g.:%
\begin{equation}
\tilde{F}_{4}^{-\mu}(k_{2},k_{1})=\hat{n}_{1}\left(  -\gamma^{\mu}\,(\hat
{v}-m)+2x_{2}v^{\mu}\right)  \,.
\end{equation}
Now we are ready to present the PJR effective vertices describing the
transition of a particle with momentum $k+q$ to a jet with momentum
$k=k_{1}+k_{2}$ and a Reggeon carrying momentum $q$ (everywhere below
$(1\leftrightarrow2)$ means in addition $G_{1}\leftrightarrow G_{2}$):%
\begin{align}
&  \Gamma_{\{G_{1}G_{2}\}Q}^{\mathcal{Q}}=g^{2}\left(  t^{G_{2}}t^{G_{1}}%
\frac{\gamma_{\bot}^{\mu}V_{1}^{\mu}(k_{2},k_{1})}{d(k_{2},k_{1})}\right.
\nonumber\\
&  -\left.  t^{G_{1}}t^{G_{2}}\frac{e_{2\bot}^{\mu}F_{3}^{\mu}(k_{2},k_{1}%
+q)}{d(k_{2},k_{1}+q)-x_{2}^{2}m^{2}}+(1\leftrightarrow2)\right)  u_{Q}\,,
\end{align}%
\begin{align}
&  \Gamma_{\{G_{1}G_{2}\}G}^{\mathcal{G}}=g^{2}2k^{-}e^{\mu}\left(
T_{G\mathcal{G}}^{a}T_{G_{1}G_{2}}^{a}\frac{V_{1}^{\mu}(k_{2},k_{1})}%
{2d(k_{2},k_{1})}\right.  \nonumber\\
&  +\left.  \left(  T^{G}T^{\mathcal{G}}\right)  _{G_{1}G_{2}}\frac{V_{1}%
^{\mu}(k_{2}+q,k_{1})}{d(k_{2}+q,k_{1})}+(1\leftrightarrow2)\right)  \,,
\end{align}
where $e^{\mu}$ is the polarization vector of the incoming gluon.%
\begin{align}
&  \Gamma_{\{Q_{1}G\}Q}^{\mathcal{G}}=-g^{2}e_{2\bot}^{\mu}\bar{u}_{Q_{1}%
}\left(  t^{G}t^{\mathcal{G}}\frac{F_{3}^{-\mu}(k\,_{2},k_{1})}{d(k\,_{2}%
,k_{1})-x_{2}^{2}m^{2}}\right.  \nonumber\\
&  +\left.  \left[  t^{\mathcal{G}}t^{G}\right]  \frac{F_{3}^{-\mu}%
(k_{2}+q,k_{1})}{d(k_{2}+q,k_{1})-x_{2}^{2}m^{2}}-t^{\mathcal{G}}t^{G}%
\frac{F_{3}^{-\mu}(k_{2},k_{1}+q)}{d(k_{2},k_{1}+q)-x_{2}^{2}m^{2}}\right)
u_{Q}\,,
\end{align}
where $k_{1}\,\,$and $k_{2}$ are momenta of the emitted quark and gluon and
$e_{2}^{\mu}$ is the polarization vector of the outgoing gluon.%
\begin{align}
&  \Gamma_{\{\bar{Q}G_{2}\}G}^{\mathcal{Q}}=-g^{2}e^{\mu}\left(  t^{G_{2}%
}t^{G}\frac{F_{4}^{\mu}(k_{2}+q,k_{1})}{d(k_{2}+q,k_{1})-m^{2}}\right.
\nonumber\\
&  +\left.  t^{G}t^{G_{2}}\frac{F_{2}^{\mu}(k\,_{2},k_{1})}{d(k\,_{2}%
,k_{1})-x_{2}^{2}m^{2}}+\left[  t^{G}t^{G_{2}}\right]  \frac{V_{3}^{\mu}%
(k_{2},k_{1}+q)}{x_{1}d(k_{2},k_{1}+q)}\right)  \upsilon_{\bar{Q}}\,,
\end{align}
where $e^{\mu}$ is the polarization vector of the incoming gluon and
$k_{1}\,\,$\ and $k_{2}$ are momenta of the emitted antiquark and gluon.%
\begin{align}
&  \Gamma_{\{Q_{1}\bar{Q}_{2}\}Q}^{\mathcal{Q}}=-\frac{g^{2}}{2k^{-}}%
t^{a}\gamma_{\bot}^{\mu}u_{Q}\otimes\bar{u}_{Q_{1}}\frac{t^{a}\tilde{F}%
_{4}^{-\mu}(k_{2},k_{1})}{d(k_{2},k_{1})-m^{2}}\upsilon_{\bar{Q}_{2}%
}\nonumber\\
&  +\frac{g^{2}}{2k_{2}^{-}}t^{a}\gamma_{\bot}^{\mu}\upsilon_{\bar{Q}_{2}%
}\otimes\bar{u}_{Q_{1}}\frac{t^{a}F_{3}^{-\mu}(k_{2}+q,k_{1})}{d(k_{2}%
+q,k_{1})-x_{2}^{2}m^{2}}u_{Q}\,,
\end{align}
where $k_{1}\,\,$and $k_{2}$ are momenta of the emitted quark and antiquark.%
\begin{align}
&  \Gamma_{\{Q_{1}\bar{Q}_{2}\}G}^{\mathcal{G}}=g^{2}e^{\mu}\bar{u}_{Q}\left(
-t^{G}t^{\mathcal{G}}\frac{\tilde{F}_{4}^{-\mu}(k_{2}+q,k_{1})}{d(k_{2}%
+q,k_{1})-m^{2}}\right.  \nonumber\\
&  +\left.  t^{\mathcal{G}}t^{G}\frac{\tilde{F}_{4}^{-\mu}(k_{2},k_{1}%
+q)}{d(k_{2},k_{1}+q)-m^{2}}+\left[  t^{G}t^{\mathcal{G}}\right]  \frac
{\tilde{F}_{4}^{-\mu}(k_{2},k_{1})}{d(k_{2},k_{1})-m^{2}}\right)
\upsilon_{\bar{Q}}\,,
\end{align}
where $e^{\mu}$ is the polarization vector of the incoming gluon and
$k_{1}\,\,$and $k_{2}$ are momenta of the emitted quark and antiquark. Quite
analogously, the RRJ effective vertices describing the production of a jet
$\{P_{1}P_{2}\}$ with momentum $k=k_{1}+k_{2}$ at the collision of two
Reggeons with momenta $-q_{1}$ and $q_{2}$ look as follows%
\begin{align}
&  \gamma_{\mathcal{G}_{1}\mathcal{G}_{2}}^{\{G_{1}G_{2}\}}%
\nonumber\label{ggtogg}\\
&  =g^{2}T_{G_{2}G_{1}}^{a}T_{\mathcal{G}_{2}\mathcal{G}_{1}}^{a}\,\left\{
\frac{V_{1}^{\mu}(k_{2},k_{1})}{d(k_{2},k_{1})}\left(  \frac{q_{2\bot}%
^{2}(k_{1}+k_{2})_{\bot}^{\mu}}{(k_{1}+k_{2})_{\bot}^{2}}-q_{2\bot}^{\mu
}\right)  +\frac{e_{1\bot}^{\mu}V_{2}^{\mu}(k_{2},k_{1})\,q_{2\bot}^{2}%
}{D(k_{2},k_{1})\,(k_{1}+k_{2})_{\bot}^{2}}\right\}  \nonumber\\
&  +g^{2}\left(  T^{G_{1}}T^{G_{2}}\right)  _{\mathcal{G}_{2}\mathcal{G}_{1}%
}\nonumber
\end{align}
\vspace{-0.9cm}
\begin{align*}
&  \,e_{1\bot}^{\mu}\left\{  -\frac{2V_{2}^{\mu}(q_{2}-k_{1},k_{1})}%
{D(q_{2}-k_{1},k_{1})\,}+q_{2\bot}^{2}\left(  \frac{{}}{{}}K_{3}^{\mu}%
(k_{2},k_{1})-K_{3}^{\mu}(q_{2}-k_{1},k_{1})\right)  \right\}  \\
&  +(1\leftrightarrow2)\,\,;
\end{align*}%
\begin{align}
&  \gamma_{\mathcal{Q}_{1}\mathcal{Q}_{2}}^{\{G_{1}G_{2}\}}=g^{2}\left[
t^{G_{1}}t^{G_{2}}\right]  \,\left\{  \left(  \frac{e_{1\bot}^{\mu}V_{2}^{\mu
}(k_{2},k_{1})}{D(k_{2},k_{1})}+\frac{(k_{1}+k_{2})_{\bot}^{\mu}V_{1}^{\mu
}(k_{2},k_{1})}{d(k_{2},k_{1})}\right)  \frac{\hat{q}_{2\bot}-m}{(k_{1}%
+k_{2})_{\bot}^{2}}\right.  \nonumber\\
&  -\left.  \frac{\gamma_{\bot}^{\mu}V_{1}^{\mu}(k_{2},k_{1})}{2d(k_{2}%
,k_{1})}\right\}  +g^{2}t^{G_{1}}t^{G_{2}}\left\{  \frac{e_{2\bot}^{\mu}%
F_{1}^{\mu}(k_{2},q_{2}-k_{2})}{D(k_{2},q_{2}-k_{2})-x_{2}m^{2}}\right.
\nonumber\\
&  +\left.  \left(  e_{2\bot}^{\mu}K_{2}^{\mu}(k_{2},q_{2}-k_{2})-e_{1\bot
}^{\mu}\tilde{K}_{3}^{\mu}(k_{2},k_{1})\right)  (\hat{q}_{2\bot}-m)\frac{{}%
}{{}}\right\}  +(1\leftrightarrow2)\,,
\end{align}
where $e_{1},e_{2}$ are the polarization vectors of the emitted gluons.%
\begin{align}
\gamma_{\mathcal{Q\,G}}^{\{\bar{Q}\,G\}} &  =g^{2}\left[  t^{G}t^{\mathcal{G}%
}\right]  \frac{\gamma_{\bot}^{\mu}}{2k_{1}^{-}}\left(  -\frac{2V_{2}^{\mu
}(k_{2},q_{2}-k_{2})}{D(k_{2},q_{2}-k_{2})}+q_{2\bot}^{2}\tilde{K}_{3}^{\mu
}(k_{2},q_{2}-k_{2})\right)  \nonumber\\
-\frac{g^{2}}{k^{-}} &  \left(  t^{\mathcal{G}}t^{G}\frac{q_{2\bot}^{\mu}%
F_{2}^{\mu}(k_{2},k_{1})}{d(k_{2},k_{1})-x_{2}^{2}m^{2}}-t^{G}t^{\mathcal{G}%
}\,\hat{e}_{2\bot}\left\{  \frac{F_{5}(k_{1},q_{2}-k_{1})}{D(q_{2}-k_{1}%
,k_{1})-m^{2}}+1\right\}  \right)  \,,
\end{align}
where $k_{1}\,$and $k_{2}$ are momenta of the emitted antiquark and gluon and
$e_{2}$ is the emitted gluon polarization;%
\begin{align*}
&  \gamma_{\mathcal{G\,Q}}^{\{Q\,G\}}\\
&  =g^{2}e_{2\bot}^{\mu}\bar{u}_{Q}\left(  \left[  t^{\mathcal{G}}%
t^{G}\right]  \,\left\{  \frac{F_{1}^{-\mu}(q_{2}-k_{1},k_{1})}{D(q_{2}%
-k_{1},k_{1})-x_{2}m^{2}}-K_{1}^{-\mu}(q_{2}-k_{1},k_{1})(\hat{q}_{2\bot
}-m)\right\}  \right.  \\
&  -t^{\mathcal{G}}t^{G}\frac{F_{1}^{-\mu}(k_{2},q_{2}-k_{2})}{D(k_{2}%
,q_{2}-k_{2})-x_{2}m^{2}}-t^{G}t^{\mathcal{G}}\,\frac{F_{3}^{-\mu}(k_{2}%
,k_{1})}{d(k_{2},k_{1})-x_{2}^{2}m^{2}}%
\end{align*}
\vspace{-0.9cm}
\begin{align}
&  +t^{\mathcal{G}}t^{G}\left(  \frac{{}}{{}}K_{1}^{-\mu}(k_{2},k_{1}%
)+K_{2}^{-\mu}(k_{2},k_{1})-K_{2}^{-\mu}(k_{2},q_{2}-k_{2})\right)  (\hat
{q}_{2\bot}-m)\nonumber\\
&  +t^{G}t^{\mathcal{G}}\,\left(  \frac{F_{3}^{-\mu}(k_{2},k_{1})}%
{d(k_{2},k_{1})-x_{2}^{2}m^{2}}+\frac{F_{1}^{-\mu}(k_{2},k_{1})}{D(k_{2}%
,k_{1})-x_{2}m^{2}}\right)  \frac{1}{\hat{k}_{1\bot}+\hat{k}_{2}{}_{\bot}%
-m}(\hat{q}_{2\bot}-m)\nonumber\\
&  -\,\left.  t^{G}t^{\mathcal{G}}K_{1}^{-\mu}(k_{2},k_{1})(\hat{q}_{2\bot
}-m)\frac{{}}{{}}\right)  ,
\end{align}
where $k_{1}\,\,$and $k_{2}$ are momenta of the emitted quark and gluon and
$e_{2}$ is the emitted gluon polarization;
\begin{align}
&  \gamma_{\mathcal{G}_{1}\mathcal{G}_{2}}^{\{Q\,\bar{Q}\}}\nonumber\\
&  =\frac{g^{2}}{k^{-}}\bar{u}_{Q}\left(  t^{\mathcal{G}_{1}}t^{\mathcal{G}%
_{2}}\frac{\hat{n}_{1}F_{5}(k_{2},q_{2}-k_{2})}{D(k_{2},q_{2}-k_{2})-m^{2}%
}-t^{\mathcal{G}_{2}}t^{\mathcal{G}_{1}}\frac{\hat{n}_{1}F_{5}(q_{2}%
-k_{1},k_{1})}{D(q_{2}-k_{1},k_{1})-m^{2}}\right)  \upsilon_{\bar{Q}}\nonumber
\end{align}
\vspace{-0.9cm}
\begin{align*}
&  +\frac{g^{2}}{k^{-}}\bar{u}_{Q}\left[  t^{\mathcal{G}_{2}}t^{\mathcal{G}%
_{1}}\right]  \frac{q_{2\bot}^{2}}{(k_{1}+k_{2})_{\bot}^{2}}\left(  \hat
{n}_{1}+\frac{\hat{n}_{1}F_{5}(k_{2},k_{1})}{D(k_{2},k_{1})-m^{2}}%
-\frac{(k_{1}+k_{2})_{\bot}^{\mu}\tilde{F}_{4}^{-\mu}(k_{2},k_{1})}%
{d(k_{2},k_{1})-m^{2}\,}\right)  \upsilon_{\bar{Q}}\\
&  +\frac{g^{2}}{k^{-}}\bar{u}_{Q}\left[  t^{\mathcal{G}_{2}}t^{\mathcal{G}%
_{1}}\right]  \left(  -\hat{n}_{1}+\frac{q_{2\bot}^{\mu}\tilde{F}_{4}^{-\mu
}(k_{2},k_{1})}{d(k_{2},k_{1})-m^{2}\,}\right)  \upsilon_{\bar{Q}}\,;
\end{align*}%
\begin{align}
&  \gamma_{\mathcal{Q}_{1}\mathcal{Q}_{2}}^{\{Q\,\bar{Q}\}}\nonumber\\
&  =-\frac{g^{2}}{2k_{2}^{-}}t^{a}\gamma_{\bot}^{\mu}\upsilon_{\bar{Q}}%
\otimes\bar{u}_{Q}t^{a}\left(  \frac{F_{1}^{-\mu}(q_{2}-k_{1},k_{1})}%
{D(q_{2}-k_{1},k_{1})-x_{2}m^{2}}-K_{1}^{\mu-}(q_{2}-k_{1},k_{1})(\hat
{q}_{2\bot}-m)\right)  \nonumber
\end{align}
\vspace{-0.9cm}
\begin{align*}
&  -g^{2}t^{a}\frac{\hat{q}_{2\bot}-m}{k^{-}(k_{1}+k_{2})_{\bot}^{2}}%
\otimes\bar{u}_{Q}t^{a}\left(  \hat{n}_{1}+\frac{\hat{n}_{1}F_{5}(k_{2}%
,k_{1})}{D(k_{2},k_{1})-m^{2}}-\frac{(k_{1}+k_{2})_{\bot}^{\mu}\tilde{F}%
_{4}^{-\mu}(k_{2},k_{1})}{d(k_{2},k_{1})-m^{2}\,}\right)  \upsilon_{\bar{Q}}\\
&  -g^{2}t^{a}\gamma_{\bot}^{\mu}\otimes\bar{u}_{Q}t^{a}\frac{\tilde{F}%
_{4}^{-\mu}(k_{2},k_{1})}{2k^{-}(d(k_{2},q_{2}-k_{2})-m^{2})}\upsilon_{\bar
{Q}}\,,
\end{align*}
where $k_{1}\,\,$and $k_{2}$ are momenta of the emitted quark and antiquark.

\section{Bootstrap conditions in QMRK}

In this article we prove a part of the quark Reggeization hypothesis in the
NLA for the QMRK or in other words the multi--Regge form
\eqref{inelastic quark} for the QMRK. Here it seems sensible to make two
remarks concerning "Reggeization" and "signaturization". These notions were
carefully discussed in \cite{Bogdan:2006af} for the case of the LLA. In the
QMRK all conclusions are valid as well with the corresponding substitution of
"jets" for "particles", so the reader is referred to the end of Section 2 of
\cite{Bogdan:2006af}.

The proof of the form (\ref{inelastic quark}) in the QMRK may be done by
obvious extension of the corresponding argumentation presented in
\cite{Bogdan:2006af} for the LLA \ It is based on the relations required by
the compatibility of the quasi--multi--Regge form of the QCD amplitudes with
the $s$--channel unitarity (bootstrap relations). Fulfillment of the bootstrap
relations impose some constraints on the Regge vertices and the trajectory
(bootstrap conditions). The bootstrap conditions are formulated in this
Section and checked in the next one.

It is worth mentioning that besides all, the argumentation in
\cite{Bogdan:2006af} uses the amplitude in Born approximation to calculate
loop--by--loop all radiative corrections to Born amplitudes and examine the
formula (\ref{inelastic quark}). Therefore, the Born form for the QMRK has to
be proved first. It is verified for the corresponding tree amplitudes from
which the vertices $\gamma_{\mathcal{RR}}^{\{P_{1}P_{2}\}}$ and $\Gamma
_{\{P_{1}^{\prime}P_{2}^{\prime}\}P}^{\mathcal{R}}$ were extracted. For the
amplitude with an arbitrary number $n$ of emitted particles the proof may be
performed via the $t$--channel unitarity as it was done in \cite{BFKL,FS} for
the MRK.

In order to present the bootstrap conditions in a compact way similar to
one in \cite{Bogdan:2006af} we use operator formalism as in
\cite{Bogdan:2006af} slightly adopting it for jet production. So,
$\langle\mathcal{G}_{i}|$ and $|\mathcal{G}_{i}\rangle$ are "bra"-- and
"ket"--vectors for $t$--channel states of the Reggeized gluon with transverse
momentum $r_{i\perp}$ and colour index $\mathcal{G}_{i}$. Their scalar product
is%
\begin{equation}
\langle\mathcal{G}_{i}|\mathcal{G}_{j}\rangle=r_{i\perp}^{2}\delta(r_{i\perp
}-r_{j\perp})\delta_{\mathcal{G}_{i}\mathcal{G}_{j}}\,. \label{norm G}%
\end{equation}
Similarly, we introduce $\langle\mathcal{Q}_{i}|$ and $|\mathcal{Q}_{i}%
\rangle$ denoting the $t$--channel states of the Reggeized quark with
transverse momentum $r_{i\perp}$, colour index $\mathcal{Q}_{i}$ and spinor
index $\rho_{i}$ and their scalar product
\begin{equation}
\langle\mathcal{Q}_{i}|\mathcal{Q}_{j}\rangle=(m-\hat{r}_{i\perp})_{\rho
_{i}\rho_{j}}\delta(r_{i\perp}-r_{j\perp})\delta_{\mathcal{Q}_{i}%
\mathcal{Q}_{j}}. \label{norm Q}%
\end{equation}
\begin{figure}[b]
\centering
\includegraphics
{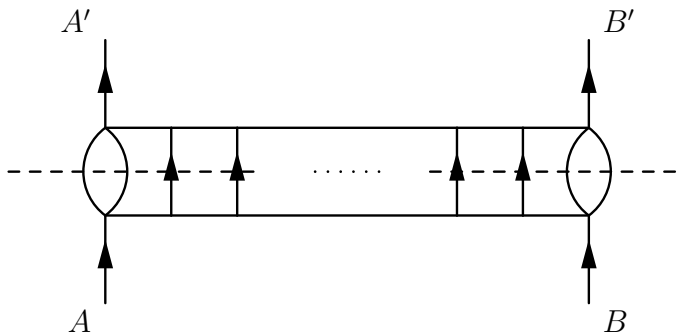} \caption{Schematic representation of
the $s$--channel discontinuity
$\mathrm{disc}_{s}^{\mathcal{S}}A_{AB\rightarrow A^{\prime}B^{\prime}}$.}%
\label{fig3}%
\end{figure}Two--Reggeon states are built from the above ones. It is useful to
distinguish the states $|\mathcal{R}_{i}\mathcal{R}_{j}\rangle$ (the
corresponding "bra"--vector $\langle\mathcal{R}_{i}\mathcal{R}_{j}|$) and
$|\mathcal{R}_{j}\mathcal{R}_{i}\rangle$ ("bra"--vector $\langle
\mathcal{R}_{j}\mathcal{R}_{i}|$). We associate the first of them with the
case when Reggeon $\mathcal{R}_{i}$ is located in the lower part of Fig.
\ref{fig3}, i.e. when it belongs to $A_{{AB}\rightarrow n+2}^{\mathcal{R}}$ in
the unitarity relation, and the second with the case when it is in the upper
part of Fig. \ref{fig3}, i.e. in the amplitude $A_{n+2\rightarrow A^{\prime
}B^{\prime}}^{\mathcal{R}}$. We define three types of states
\begin{equation}
|\mathcal{G}_{i}\mathcal{G}_{j}\rangle=|\mathcal{G}_{i}\rangle|\mathcal{G}%
_{j}\rangle,\;\;|\mathcal{G}_{i}\mathcal{Q}_{j}\rangle=|\mathcal{G}_{i}%
\rangle|\mathcal{Q}_{j}\rangle,\;\;|\mathcal{Q}_{i}\mathcal{G}_{j}%
\rangle=|\mathcal{Q}_{i}\rangle|\mathcal{G}_{j}\rangle\,.
\label{two-Reggeon states}%
\end{equation}
States of different types are orthogonal one another. All of them create a
complete set, i.e.
\begin{align}
\langle\Psi|\Phi\rangle &  =\int\langle\Psi|\mathcal{G}_{1}\mathcal{G}%
_{2}\rangle\frac{d^{D-2}r_{1\bot}d^{D-2}r_{2\bot}}{r_{1\bot}^{2}r_{2\bot}^{2}%
}\langle\mathcal{G}_{1}\mathcal{G}_{2}|\Phi\rangle\nonumber\\
&  +\int\langle\Psi|\mathcal{Q}_{1}\mathcal{G}_{2}\rangle\frac{d^{D-2}%
r_{1\bot}d^{D-2}r_{2\bot}}{(m-\hat{r}_{1\bot})r_{2\bot}^{2}}\langle
\mathcal{Q}_{1}\mathcal{G}_{2}|\Phi\rangle\nonumber\\
&  +\int\langle\Psi|\mathcal{G}_{1}\mathcal{Q}_{2}\rangle\frac{d^{D-2}%
r_{1\bot}d^{D-2}r_{2\bot}}{(m-\hat{r}_{2\bot})r_{1\bot}^{2}}\langle
\mathcal{G}_{1}\mathcal{Q}_{2}|\Phi\rangle, \label{completeness}%
\end{align}
where summation over colour and spin indices is assumed.

Bootstrap conditions relate jet production operators, impact--factors and jet
production effective vertices. We define impact--factors describing jet
production in the fragmentation regions of initial particles as the
projections of the $t$--channel states $|\{\bar{B}_{1}^{\prime}\bar{B}%
_{2}^{\prime}\}B\rangle$ or $\langle\{A_{1}^{\prime}A_{2}^{\prime}\}\,\bar
{A}|$ on two--Reggeon states (cf. eq.(32-33) in \cite{Bogdan:2006af}):
\begin{align*}
&  \langle\mathcal{R}_{1}\mathcal{R}_{2}|\{\bar{B}_{1}^{\prime}\bar{B}%
_{2}^{\prime}\}B\rangle=\delta(r_{1\perp}+r_{2\perp}-q_{B\perp})\\
&  \times\sum_{P}\left\{  \frac{1}{2p_{B}^{-}}\left(  \Gamma_{\{B_{1}^{\prime
}B_{2}^{\prime}\}P}^{\mathcal{R}_{2}}\Gamma_{PB}^{\mathcal{R}_{1}}%
\;\pm\;{\underline{\Gamma}}_{\;\bar{B}P}^{\mathcal{R}_{2}}{\underline{\Gamma}%
}_{\;P\{\bar{B}_{1}^{\prime}\bar{B}_{2}^{\prime}\}}^{\mathcal{R}_{1}}\right)
\right.
\end{align*}
\vspace{-0.9cm}
\begin{align}
&  +\frac{1}{2p_{B_{1}^{\prime}}^{-}}\left(  \Gamma_{B_{1}^{\prime}%
P}^{\mathcal{R}_{2}}\Gamma_{\{PB_{2}^{\prime}\}B}^{\mathcal{R}_{1}}%
\;\pm\;{\underline{\Gamma}}_{\;\bar{B}\{P\bar{B}_{2}^{\prime}\}}%
^{\mathcal{R}_{2}}{\underline{\Gamma}}_{\;P\bar{B}_{1}^{\prime}}%
^{\mathcal{R}_{1}}\right) \nonumber\\
&  +\left.  \frac{1}{2p_{B_{2}^{\prime}}^{-}}\left(  \Gamma_{B_{2}^{\prime}%
P}^{\mathcal{R}_{2}}\Gamma_{\{PB_{1}^{\prime}\}B}^{\mathcal{R}_{1}}%
\;\pm\;{\underline{\Gamma}}_{\;\bar{B}\{P\bar{B}_{1}^{\prime}\}}%
^{\mathcal{R}_{2}}{\underline{\Gamma}}_{\;P\bar{B}_{2}^{\prime}}%
^{\mathcal{R}_{1}}\right)  \right\}  , \label{impact b-rr}%
\end{align}
where $\Gamma_{\{B_{1}^{\prime}B_{2}^{\prime}\}B}^{\mathcal{R}}$ are the
particle--jet--Reggeon (PJR) effective vertices describing particle--jet
$P\rightarrow\{P_{1}^{\prime}P_{2}^{\prime}\}$ transition due to the
interaction with Reggeon $\mathcal{R}$; the $+$ ($-$) sign stands for the
fermion (boson) state in the $t$--channel, $q_{B}=p_{B}-p_{B_{1}^{\prime}%
}-p_{B_{2}^{\prime}}$, the sum is taken over quantum numbers of particles $P$
(they can be different in different terms) and the factor $1/p_{B_{i}}^{-}$
comes from the phase space element in the unitarity relation (see eqs. (27,38)
in \cite{Bogdan:2006af}). The factor $1/2$ and the last term in each brackets
in \eqref{impact
b-rr} stand on account of the "signaturization". The bar over particle symbol
means, as usual, antiparticle while ${\underline{\Gamma}}_{\;\bar{B}%
P}^{\mathcal{R}_{2}}\,$, ${\underline{\Gamma}}_{\;\bar{B}\{P\bar{B}%
_{i}^{\prime}\}}^{\mathcal{R}_{2}}$ and ${\underline{\Gamma}}_{\;P\bar
{B}^{\prime}}^{\mathcal{R}_{1}}\,$, ${\underline{\Gamma}}_{\;P\{\bar{B}%
_{1}^{\prime}\bar{B}_{2}^{\prime}\}}^{\mathcal{R}_{1}}$ are obtained from
${\Gamma}_{\bar{B}P}^{\mathcal{R}_{2}}\,$, ${\Gamma}_{\;\bar{B}\{P\bar{B}%
_{i}^{\prime}\}}^{\mathcal{R}_{2}}$ and ${\Gamma}_{P\bar{B}^{\prime}%
}^{\mathcal{R}_{1}}\,$, ${\Gamma}_{\;P\{\bar{B}_{1}^{\prime}\bar{B}%
_{2}^{\prime}\}}^{\mathcal{R}_{1}}$ correspondingly taking instead of wave
functions (polarization vectors and Dirac spinors) of $\bar{B}$ and $\bar
{B}_{i}^{\prime}$ the wave functions of $B$ and $B_{i}^{\prime}$ from the
first term in brackets of \eqref{impact
b-rr}.

Quite analogously,
\begin{align*}
&  \langle\{{A_{1}^{\prime}A}_{2}^{\prime}\}{\bar{A}|\mathcal{R}}%
_{1}\mathcal{R}_{2}\rangle=\delta(r_{1\perp}+r_{2\perp}-q_{A\perp})\\
&  \times\sum_{P}\left\{  \frac{1}{2p_{A}^{+}}\left(  \bar{\Gamma}%
_{\{A_{1}^{\prime}A_{2}^{\prime}\}P}^{\mathcal{R}_{2}}\bar{\Gamma}%
_{PA}^{\mathcal{R}_{1}}\;\pm\;{\underline{\bar{\Gamma}}}_{\;\bar{A}%
P}^{\mathcal{R}_{2}}{\underline{\bar{\Gamma}}}_{\;P\{\bar{A}_{1}^{\prime}%
\bar{A}_{2}^{\prime}\}}^{\mathcal{R}_{1}}\right)  \right.
\end{align*}
\vspace{-0.9cm}
\begin{align}
&  +\frac{1}{2p_{A_{1}^{\prime}}^{+}}\left(  \bar{\Gamma}_{A_{1}^{\prime}%
P}^{\mathcal{R}_{2}}\bar{\Gamma}_{\{PA_{2}^{\prime}\}A}^{\mathcal{R}_{1}}%
\;\pm\;{\underline{\bar{\Gamma}}}_{\;\bar{A}\{P\bar{A}_{2}^{\prime}%
\}}^{\mathcal{R}_{2}}{\underline{\bar{\Gamma}}}_{\;P\bar{A}_{1}^{\prime}%
}^{\mathcal{R}_{1}}\right) \nonumber\\
&  \left.  \frac{1}{2p_{A_{2}^{\prime}}^{+}}\left(  \bar{\Gamma}%
_{A_{2}^{\prime}P}^{\mathcal{R}_{2}}\bar{\Gamma}_{\{PA_{1}^{\prime}%
\}A}^{\mathcal{R}_{1}}\;\pm\;{\underline{\bar{\Gamma}}}_{\;\bar{A}\{P\bar
{A}_{1}^{\prime}\}}^{\mathcal{R}_{2}}{\underline{\bar{\Gamma}}}_{\;P\bar
{A}_{2}^{\prime}}^{\mathcal{R}_{1}}\right)  \right\}  , \label{impact a-rr}%
\end{align}
where $q_{A}=p_{A_{1}^{\prime}}+p_{A_{2}^{\prime}}-p_{A}$.

The strong bootstrap conditions resulting from the quasi--elastic (one final
particle is a two--particle jet) amplitudes impose the following constraints
on the impact factors%
\begin{equation}
|\{{\bar{B}_{1}^{\prime}\bar{B}}_{2}^{\prime}\}{B}\rangle=g|{\mathcal{R}%
}_{\omega}(q_{B\bot})\rangle\Gamma_{\{B_{1}^{\prime}B_{2}^{\prime}%
\}B}^{\mathcal{R}},\,\langle\{{A_{1}^{\prime}A}_{2}^{\prime}\}{\bar{A}}%
|=g\bar{\Gamma}_{\{A_{1}^{\prime}A_{2}^{\prime}\}A}^{\mathcal{R}}%
\langle{\mathcal{R}}_{\omega}(q_{A\bot})|\,,\label{bootstrap vertex}%
\end{equation}
where $|\mathcal{R}_{\omega}(q_{\bot})\rangle$ are universal (process
independent) eigenstates of the kernel ${\hat{\mathcal{K}}}$ (see eq.(43) in
\cite{Bogdan:2006af}) with the eigenvalues ${\omega_{\mathcal{R}}(q)}$. From
calculations in leading order we know that
\begin{align}
\langle\mathcal{G}_{1}\mathcal{G}_{2}|\mathcal{G}_{\omega}(q_{\bot})\rangle &
=\delta(r_{1\perp}+r_{2\perp}-q_{\bot}){T}_{\mathcal{G}_{1}\mathcal{G}_{2}%
}^{\mathcal{G}}\,,\nonumber\\
\langle\mathcal{G}_{1}\mathcal{Q}_{2}|\mathcal{Q}_{\omega}(q_{\bot})\rangle &
=\delta(r_{1\perp}+r_{2\perp}-q_{\bot}){t}^{\mathcal{G}_{1}},\nonumber\\
\langle\mathcal{Q}_{1}\mathcal{G}_{2}|\mathcal{Q}_{\omega}(q_{\bot})\rangle &
=-\delta(r_{1\perp}+r_{2\perp}-q_{\bot}){t}^{\mathcal{G}_{2}}%
.\label{impact G-gg}%
\end{align}

Similarly to the impact factors for scattering jets we define the impact
factors for Reggeon--jet transitions (compare with \eqref{impact a-rr}) as
\begin{align*}
&  \langle\mathcal{R}_{1}\mathcal{R}_{2}|\{\bar{P}_{1}\bar{P}_{2}%
\}\mathcal{R}_{j+1}\rangle=\delta(r_{1\perp}+r_{2\perp}-q_{j\bot})\\
&  \times\sum_{P}\left\{  \frac{1}{2k^{-}}\left(  \Gamma_{\{P_{1}P_{2}%
\}P}^{\mathcal{R}_{2}}\gamma_{\mathcal{R}_{1}\mathcal{R}_{j+1}}^{P}%
\;\pm\;{\underline{\Gamma}}_{\;P\{\bar{P}_{1}\bar{P}_{2}\}}^{\mathcal{R}_{1}%
}\gamma_{P}^{\mathcal{R}_{2}\mathcal{R}_{j+1}}\right)  \right.
\end{align*}
\vspace{-0.9cm}
\begin{align}
&  +\frac{1}{2k_{1}^{-}}\left(  \Gamma_{P_{1}P}^{\mathcal{R}_{2}}%
\gamma_{\mathcal{R}_{1}\mathcal{R}_{j+1}}^{\{PP_{2}\}}\;\pm\;{\underline
{\Gamma}}_{\;P\bar{P}_{1}}^{\mathcal{R}_{1}}\gamma_{\{P\bar{P}_{2}%
\}}^{\mathcal{R}_{2}\mathcal{R}_{j+1}}\right) \nonumber\\
&  +\left.  \frac{1}{2k_{2}^{-}}\left(  \Gamma_{P_{2}P}^{\mathcal{R}_{2}%
}\gamma_{\mathcal{R}_{1}\mathcal{R}_{j+1}}^{\{PP_{1}\}}\;\pm\;{\underline
{\Gamma}}_{\;P\bar{P}_{2}}^{\mathcal{R}_{1}}\gamma_{\{P\bar{P}_{1}%
\}}^{\mathcal{R}_{2}\mathcal{R}_{j+1}}\right)  \right\}  \,, \label{impact rp}%
\end{align}
where $\gamma_{\mathcal{R}_{i}\mathcal{R}_{i+1}}^{\{P_{1}P_{2}\}}$ are the RRJ
effective vertices, describing production of jets $\left\{  P_{1}%
P_{2}\right\}  $ at the Reggeon transition $\mathcal{R}_{i+1}\rightarrow
\mathcal{R}_{i}$; $k=k_{1}+k_{2}$ is the jet momentum, $q_{j\bot}%
=q_{(j+1)\bot}-k_{\bot}$, the $+(-)$ sign stands for the case when the Reggeon
quantum numbers (i.e. quark or gluon) in the $j$ and $j+1$ channels are equal
(different). Analogously%
\begin{align*}
&  \langle\{{P_{1}P}_{2}\}{\mathcal{R}_{i}|\mathcal{R}}_{1}\mathcal{R}%
_{2}\rangle=\delta(r_{1\perp}+r_{2\perp}-q_{(i+1)\bot})\\
&  \times\sum_{P}\left\{  \frac{1}{2k^{+}}\left(  \bar{\Gamma}_{\{P_{1}%
P_{2}\}P}^{\mathcal{R}_{2}}\gamma_{\mathcal{R}_{i}\mathcal{R}_{1}}^{P}%
\;\pm\;{\underline{\bar{\Gamma}}}_{\;P\{\bar{P}_{1}\bar{P}_{2}\}}%
^{\mathcal{R}_{1}}\gamma_{P}^{\mathcal{R}_{i}\mathcal{R}_{2}}\right)  \right.
\end{align*}
\vspace{-0.9cm}
\begin{align}
&  +\frac{1}{2k_{1}^{+}}\left(  \bar{\Gamma}_{P_{1}P}^{\mathcal{R}_{2}}%
\gamma_{\mathcal{R}_{i}\mathcal{R}_{1}}^{\{PP_{2}\}}\;\pm\;{\underline
{\bar{\Gamma}}}_{\;P\bar{P}_{1}}^{\mathcal{R}_{1}}\gamma_{\{P\bar{P}_{2}%
\}}^{\mathcal{R}_{i}\mathcal{R}_{2}}\right) \nonumber\\
&  +\left.  \frac{1}{2k_{2}^{+}}\left(  \bar{\Gamma}_{P_{2}P}^{\mathcal{R}%
_{2}}\gamma_{\mathcal{R}_{i}\mathcal{R}_{1}}^{\{PP_{1}\}}\;\pm\;{\underline
{\bar{\Gamma}}}_{\;P\bar{P}_{2}}^{\mathcal{R}_{1}}\gamma_{\{P\bar{P}_{1}%
\}}^{\mathcal{R}_{i}\mathcal{R}_{2}}\right)  \right\}  , \label{impact pr}%
\end{align}
where $q_{(i+1)\bot}=q_{i\bot}+k_{\bot}$.

In expressions (\ref{impact rp}),(\ref{impact pr}) we introduced the effective
vertices $\gamma_{\{\bar{P}_{1}\bar{P}_{2}\}}^{\mathcal{R}_{i}\mathcal{R}_{j}%
}.$ One can obtain them from $\gamma_{\mathcal{R}_{i}\mathcal{R}_{j}}%
^{\{P_{1}P_{2}\}}$ replacing the wave--functions of the emitted particles with
the wave--functions of the corresponding incoming antiparticles:%
\begin{equation}
\bar{u}_{Q}\rightarrow\bar{\upsilon}_{\bar{Q}},\quad\upsilon_{\bar{Q}%
}\rightarrow u_{Q},\quad e_{G}^{\ast}\rightarrow e_{G}%
\end{equation}
and inverting the momenta $k_{P_{i}}\rightarrow-k_{\bar{P}_{i}}.$ In fermion
case we also have to change the overall sign due to the operator ordering.
There is no uniformity in literature on the matter of including this factor
$(-1)$ into the definition of the corresponding effective vertices
$\gamma_{\{\bar{P}_{1}\bar{P}_{2}\}}^{\mathcal{R}_{i}\mathcal{R}_{j}}$ or into
the impact--factor definition. We define all pair--production vertices without
it. As for the RRJ effective vertices, we follow the denotations from
\cite{Bogdan:2006af}, where $\gamma_{\bar{Q}}^{\mathcal{GQ}}$ is defined with
$(-1)$. Thus, when this factor arises and it is not included in the RRJ or PPR
effective vertex we explicitly write it in the impact--factor.

Finally, we introduce the operator $\widehat{\{{P}_{1}P_{2}\}}$ for the
production of a jet $\left\{  P_{1}P_{2}\right\}  $ with the overall momentum
$k=k_{1}+k_{2}$ as having the following matrix elements:
\begin{align}
&  \langle\mathcal{R}_{1}\mathcal{R}_{2}|\widehat{\{{P}_{1}P_{2}%
\}}|\mathcal{R}_{1}^{\prime}\mathcal{R}_{2}^{\prime}\rangle=\delta
(q_{(l+1)\perp}-k_{\perp}-q_{l\perp})\nonumber\\
&  \left(  \gamma_{\mathcal{R}_{1}\mathcal{R}_{1}^{\prime}}^{\{P_{1}P_{2}%
\}}\delta_{\mathcal{R}_{2}\mathcal{R}_{2}^{\prime}}\delta(r_{2\bot}-r_{2\bot
}^{\prime})d_{R_{2}}+\gamma_{\mathcal{R}_{2}\mathcal{R}_{2}^{\prime}}%
^{\{P_{1}P_{2}\}}\delta_{\mathcal{R}_{1}\mathcal{R}_{1}^{\prime}}%
\delta(r_{1\bot}-r_{1\bot}^{\prime})d_{\mathcal{R}_{1}}\right. \nonumber\\
&  +\left.  \gamma_{\mathcal{R}_{1}\mathcal{R}_{1}^{\prime}}^{P_{1}}%
\gamma_{\mathcal{R}_{2}\mathcal{R}_{2}^{\prime}}^{P_{2}}\delta(k_{1}+r_{1\bot
}-r_{1\bot}^{\prime})+\gamma_{\mathcal{R}_{1}\mathcal{R}_{1}^{\prime}}^{P_{2}%
}\gamma_{\mathcal{R}_{2}\mathcal{R}_{2}^{\prime}}^{P_{1}}\delta(k_{2}%
+r_{1\bot}-r_{1\bot}^{\prime})\right)  \,, \label{production operator}%
\end{align}
where $q_{l\perp}=r_{1\perp}+r_{2\perp},\;\;q_{(l+1)\perp}=r_{1\perp}^{\prime
}+r_{2\perp}^{\prime}$.

An additional bootstrap condition, which may be obtained from the bootstrap
relation for the amplitudes of a process $A+B\rightarrow A^{\prime}%
+\{P_{1}P_{2}\}+B^{\prime}$, is analogous to the LLA one \cite{Bogdan:2006af}.
For "ket"--vectors it reads as follows
\begin{equation}
\widehat{\{{P}_{1}P_{2}\}}|\mathcal{R}_{\omega}(q_{(i+1)\bot})\rangle
gd_{i+1}(q_{(i+1)\bot})+|\{\bar{P}_{1}\bar{P}_{2}\}\mathcal{R}_{i+1}%
\rangle=|\mathcal{R}_{\omega}(q_{i\bot})\rangle g\gamma_{\mathcal{R}%
_{i}\mathcal{R}_{i+1}}^{\{P_{1}P_{2}\}}, \label{bootstrap production ket}%
\end{equation}
where $q_{i\bot}=q_{(i+1)\bot}-k_{\bot},$ while for "bra"--vectors this
condition has the form
\begin{equation}
g\,\,d_{i}(q_{i\bot})\,\langle\mathcal{R}_{\omega}(q_{i\bot})|\widehat
{\{{P}_{1}P_{2}\}}\,+\,\langle\{P_{1}P_{2}\}\mathcal{R}_{i}|\,=g\,\,\gamma
_{\mathcal{R}_{i}\mathcal{R}_{i+1}}^{\{P_{1}P_{2}\}}\,\langle\mathcal{R}%
_{\omega}(q_{(i+1)\bot})|\,, \label{bootstrap production bra}%
\end{equation}
where $q_{(i+1)\bot}=q_{i\bot}+k_{\bot}$.

Note that the conditions for "ket"-- and "bra"-- vectors in
(\ref{bootstrap production ket}--\ref{bootstrap production bra}) and
(\ref{bootstrap vertex}) are not independent since these vectors are
interrelated. Indeed, the replacement of "$+$" and "$-$" momenta components
turns any of them into the other, so we consider only "ket"--vectors herein.

Jet production in the central region in the Reggeized gluon collision and in
the fragmentation region with the Reggeized gluon \ was considered in
\cite{Fadin:2003xs}. We investigate here the bootstrap conditions for
Reggeized quarks.

\section{Verification of bootstrap conditions on Reggeon vertices}

In this Section we explicitly show that bootstrap conditions
(\ref{bootstrap vertex}) and (\ref{bootstrap production ket}%
)--(\ref{bootstrap production bra}) are satisfied by the known expressions for
the effective vertices presented in Section 2. For this purpose we have chosen
such a spacial parametrization of these Regge vertices that their following
insertion into the bootstrap conditions leads to trivial cancellations.

Each bootstrap condition on concrete Regge vertex we check only for
$\langle\mathcal{G}_{1}\mathcal{Q}_{2}|$ (or for $\langle\mathcal{Q}%
_{1}\mathcal{G}_{2}|$). Looking at the diagrams one can see that in case of
$\langle\mathcal{Q}_{1}\mathcal{G}_{2}|$ ($\langle\mathcal{G}_{1}%
\mathcal{Q}_{2}|$) the calculation is quite analogous being different only in
an overall "$-$" sign and the replacement $r_{2}\longleftrightarrow r_{1}$.

\subsection{Two--gluon jet production in fragmentation region}

We begin with two--gluon jet production in the fragmentation region. In this
and the next subsection let us denote the momentum of the incoming quark
$Q_{B}$ as $p_{B}$ and the momenta of the outgoing Reggeized gluon
$\mathcal{G}_{1}$ and quark $\mathcal{Q}_{2}$ as $\,r_{1}$ and $r_{2}$
respectively, $r_{1}+r_{2}=q$. Here the momenta of the emitted gluons are
$k_{1}$ and $k_{2}$. We also often meet shifted momenta%
\begin{equation}
k_{1\bot}^{\prime}=k_{1\bot}+r_{1\bot},\quad k_{2\bot}^{\prime}=k_{2\bot
}+r_{2\bot},\quad k_{2\bot}^{\prime\prime}=k_{2\bot}+r_{1\bot},\quad k_{1\bot
}^{\prime\prime}=k_{1\bot}+r_{2\bot}.
\end{equation}
\smallskip\ The bootstrap condition for\textbf{ }$\Gamma_{\{G_{1}G_{2}%
\}Q}^{\mathcal{Q}}$ has the form:%
\begin{equation}
\langle\mathcal{G}_{1}\mathcal{Q}_{2}|\{\bar{G}_{1}\bar{G}_{2}\}Q_{B}%
\rangle=\langle\mathcal{G}_{1}\mathcal{Q}_{2}|\mathcal{Q}_{\omega}(q_{\bot
})\rangle\mathcal{\,}g\,\Gamma_{\{G_{1}G_{2}\}Q_{B}}^{\mathcal{Q}%
},\label{bootstrap_GG_frag}%
\end{equation}
where
\begin{align*}
&  \langle\mathcal{G}_{1}\mathcal{Q}_{2}|\{\bar{G}_{1}\bar{G}_{2}%
\}Q_{B}\rangle=\delta(q_{\bot}+k_{\bot}-p_{B\bot})\\
&  \times\left[  \frac{1}{4k^{-}}\left(  \sum_{Q}\Gamma_{\{G_{1}G_{2}%
\}Q}^{\mathcal{Q}_{2}}\,\Gamma_{QQ_{B}}^{\mathcal{G}_{1}}+\sum_{G}%
\underline{\Gamma}_{G\{\bar{G}_{1}\bar{G}_{2}\}}^{\mathcal{G}_{1}}%
\,\underline{\Gamma}_{\bar{Q}_{B}G}^{\mathcal{Q}_{1}}\right)  \right.
\end{align*}
\vspace{-0.9cm}
\begin{align}
&  +\frac{1}{2k_{1}^{-}}\left(  \sum_{Q}\Gamma_{G_{1}Q}^{\mathcal{Q}_{2}%
}\Gamma_{\{G_{2}Q\}Q_{B}}^{\mathcal{G}_{1}}+\sum_{G}\underline{\Gamma
}_{G\,\bar{G}_{1}}^{\mathcal{G}_{1}}\underline{\Gamma}_{\bar{Q}\,_{B}%
\{G\,\bar{G}_{2}\}}^{\mathcal{Q}_{2}}\right)  \nonumber\\
&  +1\leftrightarrow2\label{GGimp_fact}%
\end{align}
As in the definition of the effective vertices we use the denotation
($1\leftrightarrow2)$ assuming the replacement $\{e_{1\bot},k_{1\bot}%
,x_{1},G_{1}\}\leftrightarrow\{e_{2\bot},k_{2\bot},x_{2},G_{2}\}$. The parts
of (\ref{GGimp_fact}) yield:%
\begin{align}
&  \frac{1}{2k^{-}}\sum_{Q}\Gamma_{\{G_{1}G_{2}\}Q}^{\mathcal{Q}_{2}}%
\,\Gamma_{QQ_{B}}^{\mathcal{G}_{1}}=g^{3}\left(  -t^{G_{2}}t^{G_{1}%
}t^{\mathcal{G}_{1}}\frac{e_{1\bot}^{\mu}\tilde{F}_{3}^{\mu}(k_{2}^{\prime
},k_{1})}{d(k_{2}^{\prime},k_{1})-x_{1}^{2}m^{2}}\right.  \nonumber\\
&  +\left.  t^{G_{2}}t^{G_{1}}t^{\mathcal{G}_{1}}\frac{\gamma_{\bot}^{\mu
}V_{1}^{\mu}(k_{2},k_{1})}{d(k_{2},k_{1})}+(1\leftrightarrow2)\right)
u_{Q_{B}},
\end{align}%
\begin{align}
&  \frac{1}{2k^{-}}\sum_{G}\underline{\Gamma}_{G\{\bar{G}_{1}\bar{G}_{2}%
\}}^{\mathcal{G}_{1}}\,\underline{\Gamma}_{\bar{Q}_{B}G}^{\mathcal{Q}_{1}%
}=-g^{3}\left(  \left[  t^{G_{2}}t^{G_{1}},t^{\mathcal{G}_{1}}\right]
\frac{V_{1}^{\mu}(k_{2},k_{1})}{d(k_{2},k_{1})}\right.  \nonumber\\
&  +\left.  \left[  t^{G_{1}}\left[  t^{G_{2}}t^{\mathcal{G}_{1}}\right]
\right]  \frac{V_{1}^{\mu}(k_{2}^{\prime\prime},k_{1})}{d(k_{2}^{\prime\prime
},k_{1})}+(1\leftrightarrow2)\right)  \gamma_{\bot}^{\mu}u_{Q_{B}},
\end{align}%
\begin{align}
&  \frac{1}{2k_{1}^{-}}\sum_{Q}\Gamma_{G_{1}Q}^{\mathcal{Q}_{2}}%
\Gamma_{\{G_{2}Q\}Q_{B}}^{\mathcal{G}_{1}}=g^{3}t^{G_{1}}\left(  t^{G_{2}%
}t^{\mathcal{G}_{1}}\frac{e_{2\bot}^{\mu}F_{3}^{\mu}(k\,_{2},k_{1}%
^{\prime\prime})}{d(k\,_{2},k_{1}^{\prime\prime})-x_{2}^{2}m^{2}}\right.
\nonumber\\
&  +\left.  \left[  t^{\mathcal{G}_{1}}t^{G_{2}}\right]  \frac{e_{2\bot}^{\mu
}F_{3}^{\mu}(k_{2}^{\prime\prime},k_{1}^{\prime\prime})}{d(k_{2}^{\prime
\prime},k_{1}^{\prime\prime})-x_{2}^{2}m^{2}}-t^{\mathcal{G}_{1}}t^{G_{2}%
}\frac{e_{2\bot}^{\mu}F_{3}^{\mu}(k_{2},p_{B}-k_{2})}{d(k_{2},p_{B}%
-k_{2})-x_{2}^{2}m^{2}}\right)  u_{Q_{B}},\label{commutation_need}%
\end{align}%
\begin{align}
&  \frac{1}{2k_{1}^{-}}\sum_{G}\underline{\Gamma}_{G\,\bar{G}_{1}%
}^{\mathcal{G}_{1}}\underline{\Gamma}_{\bar{Q}\,_{B}\{G\,\bar{G}_{2}%
\}}^{\mathcal{Q}_{2}}=g^{3}\left(  -t^{G_{2}}\left[  t^{\mathcal{G}_{1}%
}t^{G_{1}}\right]  \frac{e_{1\bot}^{\mu}\tilde{F}_{3}^{\mu}(k_{2}^{\prime
},k_{1}^{\prime})}{d(k_{2}^{\prime},k_{1}^{\prime})-x_{1}^{2}m^{2}}\right.
\nonumber\\
&  -\left.  \left[  t^{\mathcal{G}_{1}}t^{G_{1}}\right]  t^{G_{2}}%
\frac{e_{2\bot}^{\mu}F_{3}^{\mu}(k_{2},p_{B}-k_{2})}{d(k_{2},p_{B}%
-k_{2})-x_{2}^{2}m^{2}}+\left[  t^{G_{2}}\left[  t^{\mathcal{G}_{1}}t^{G_{1}%
}\right]  \right]  \frac{\gamma_{\bot}^{\mu}V_{1}^{\mu}(k_{2},k_{1}^{\prime}%
)}{d(k_{2},k_{1}^{\prime})}\right)  u_{Q_{B}}.\label{G1G2emission4}%
\end{align}
Here we use the relation $\hat{e}_{1\bot}\hat{n}_{2}F_{3}^{-\mu}=2F_{3}^{\mu
}-\hat{n}_{1}(\dots)$ to obtain (\ref{commutation_need}). Now one can clearly
see that (\ref{bootstrap_GG_frag}) is an identity.

The bootstrap condition for antiquark--gluon production reads as%

\begin{equation}
\langle\mathcal{G}_{1}\mathcal{Q}_{2}|\{Q\bar{G}\}G_{B}\rangle=\langle
\mathcal{G}_{1}\mathcal{Q}_{2}|\mathcal{Q}_{\omega}(q_{\bot})\rangle
\mathcal{\,}g\,\Gamma_{\{\bar{Q}G\}G_{B}}^{\mathcal{Q}}\,,
\end{equation}
\qquad\qquad

but we need not check it because $\Gamma_{\{\bar{Q}G\}G_{B}}^{\mathcal{Q}}$ is
connected with the two--gluon production effective vertex $\Gamma
_{\{G_{1}G_{2}\}Q_{B}}^{\mathcal{Q}}$ by crossing rules.

\subsection{Quark--antiquark jet production in fragmentation region}

We denote the momenta of the emitted quark $Q_{1}$ and antiquark $\bar{Q}_{2}$
as $k_{1}$ and $k_{2}$. \smallskip The bootstrap condition for\textbf{
}$\Gamma_{\{Q_{1}\bar{Q}_{2}\}Q}^{\mathcal{Q}}$ has the form:%
\begin{equation}
\langle\mathcal{G}_{1}\mathcal{Q}_{2}|\{\bar{Q}_{1}Q_{2}\}Q_{B}\rangle
=\langle\mathcal{G}_{1}\mathcal{Q}_{2}|\mathcal{Q}_{\omega}(q_{\bot}%
)\rangle\mathcal{\,}g\,\Gamma_{\{Q_{1}\bar{Q}_{2}\}Q}^{\mathcal{Q}%
},\label{bootstrap_QbarQ_frag}%
\end{equation}
where
\begin{align}
&  \langle\mathcal{G}_{1}\mathcal{Q}_{2}|\{\bar{Q}_{1}Q_{2}\}Q_{B}%
\rangle=\delta(q_{\bot}+k_{\bot}-p_{B\bot})\nonumber\\
&  \times\left[  \frac{1}{2k^{-}}\left(  \sum_{Q}\Gamma_{\{Q_{1}\bar{Q}%
_{2}\}Q}^{\mathcal{Q}_{2}}\,\Gamma_{QQ_{B}}^{\mathcal{G}_{1}}+\sum
_{G}\underline{\Gamma}_{G\{\bar{Q}_{1}Q_{2}\}}^{\mathcal{G}_{1}}%
\,\underline{\Gamma}_{\bar{Q}_{B}G}^{\mathcal{Q}_{2}}\right)  +\frac{1}%
{2k_{1}^{-}}\sum_{\bar{Q}}\underline{\Gamma}_{\bar{Q}\bar{Q}_{1}}%
^{\mathcal{G}_{1}}\underline{\Gamma}\,_{\bar{Q}_{B}\{\bar{Q}Q_{2}%
\}}^{\mathcal{Q}_{2}}\right.  \nonumber\\
&  +\left.  \frac{1}{2k_{2}^{-}}\left(  \sum_{G}\Gamma_{\bar{Q}_{2}%
G}^{\mathcal{Q}_{2}}\Gamma_{\{GQ_{1}\}Q_{B}}^{\mathcal{G}_{1}}+\sum
_{Q}\underline{\Gamma}_{Q\,Q_{2}}^{\mathcal{G}_{1}}\underline{\Gamma}_{\bar
{Q}\,_{B}\{\bar{Q}_{1}Q\}}^{\mathcal{Q}_{2}}\right)  \right]
\,.\label{QbarQimp_fact}%
\end{align}
The parts of (\ref{QbarQimp_fact}) yield:%
\begin{align}
&  \frac{1}{2k^{-}}\sum_{Q}\Gamma_{\{Q_{1}\bar{Q}_{2}\}Q}^{\mathcal{Q}_{2}%
}\,\Gamma_{QQ_{B}}^{\mathcal{G}_{1}}=\frac{g^{3}}{2k^{-}}\left[
-\,t^{a}t^{\mathcal{G}_{1}}\gamma_{\bot}^{\mu}u_{Q_{B}}\otimes\bar{u}_{Q_{1}%
}\frac{t^{a}\tilde{F}_{4}^{-\mu}(k_{2},k_{1})}{d(k_{2},k_{1})-m^{2}}%
\upsilon_{\bar{Q}_{2}}\right.  \nonumber\\
&  \left.  +\frac{t^{a}\gamma_{\bot}^{\mu}}{x_{2}}\upsilon_{\bar{Q}_{2}%
}\otimes\bar{u}_{Q_{1}}\frac{t^{a}t^{\mathcal{G}_{1}}F_{3}^{-\mu}%
(k_{2}^{\prime},k_{1})}{d(k_{2}^{\prime},k_{1})-x_{2}^{2}m^{2}}u_{Q_{B}%
}\right]  ,
\end{align}%
\begin{align}
&  \frac{1}{2k^{-}}\sum_{G}\underline{\Gamma}_{G\{\bar{Q}_{1}Q_{2}%
\}}^{\mathcal{G}_{1}}\,\underline{\Gamma}_{\bar{Q}_{B}G}^{\mathcal{Q}_{2}%
}=-\frac{g^{3}}{2k^{-}}t^{a}\gamma_{\bot}^{\mu}u_{Q_{B}}\otimes\bar{u}%
_{Q}\left[  -t^{a}t^{\mathcal{G}_{1}}\frac{\tilde{F}_{4}^{-\mu}(k_{2}%
^{\prime\prime},k_{1})}{d(k_{2}^{\prime\prime},k_{1})-m^{2}}\right.
\nonumber\\
&  +\left.  t^{\mathcal{G}_{1}}t^{a}\frac{\tilde{F}_{4}^{-\mu}(k_{2}%
,k_{1}^{\prime})}{d(k_{2},k_{1}^{\prime})-m^{2}}+\left[  t^{a}t^{\mathcal{G}%
_{1}}\right]  \frac{\tilde{F}_{4}^{-\mu}(k_{2},k_{1})}{d(k_{2},k_{1})-m^{2}%
}\right]  \upsilon_{\bar{Q}_{2}},
\end{align}%
\begin{align}
&  \frac{1}{2k_{1}^{-}}\sum_{\bar{Q}}\underline{\Gamma}_{\bar{Q}\bar{Q}_{1}%
}^{\mathcal{G}_{1}}\underline{\Gamma}\,_{\bar{Q}_{B}\{\bar{Q}Q_{2}%
\}}^{\mathcal{Q}_{2}}=-\frac{g^{3}}{2k^{-}}\left[  -t^{a}\gamma_{\bot}^{\mu
}u_{Q_{B}}\otimes\bar{u}_{Q_{1}}\frac{t^{\mathcal{G}_{1}}t^{a}\tilde{F}%
_{4}^{-\mu}(k_{2},k_{1}^{\prime})}{d(k_{2},k_{1}^{\prime})-m^{2}}%
\upsilon_{\bar{Q}_{2}}\right.  \nonumber\\
&  +\left.  \frac{t^{a}\gamma_{\bot}^{\mu}}{x_{2}}\upsilon_{\bar{Q}_{2}%
}\otimes\bar{u}_{Q_{1}}\frac{t^{\mathcal{G}_{1}}t^{a}F_{3}^{-\mu}%
(k_{2}^{\prime},k_{1}^{\prime})}{d(k_{2}^{\prime},k_{1}^{\prime})-x_{2}%
^{2}m^{2}}u_{Q_{B}}\right]  \,,
\end{align}%
\begin{align}
&  \frac{1}{2k_{2}^{-}}\sum_{G}\Gamma_{\bar{Q}_{2}G}^{\mathcal{Q}_{2}}%
\Gamma_{\{GQ_{1}\}Q_{B}}^{\mathcal{G}_{1}}=-\frac{g^{3}}{2x_{2}k^{-}}%
t^{a}\gamma_{\bot}^{\mu}\upsilon_{\bar{Q}_{2}}\otimes\bar{u}_{Q_{1}}\left[
t^{a}t^{\mathcal{G}_{1}}\frac{F_{3}^{-\mu}(k^{\prime}{}_{2},k_{1}%
)}{d(k^{\prime}{}_{2},k_{1})-x_{2}^{2}m^{2}}\right.  \nonumber\\
&  +\left.  \left[  t^{\mathcal{G}_{1}}t^{a}\right]  \frac{F_{3}^{-\mu}%
(p_{B}-k_{1},k_{1})}{d(p_{B}-k_{1},k_{1})-x_{2}^{2}m^{2}}-t^{\mathcal{G}_{1}%
}t^{a}\frac{F_{3}^{-\mu}(k_{2}^{\prime},k_{1}^{\prime})}{d(k_{2}^{\prime
},k_{1}^{\prime})-x_{2}^{2}m^{2}}\right]  u_{Q}\,,
\end{align}%
\begin{align}
&  \frac{1}{2k_{2}^{-}}\sum_{Q}\underline{\Gamma}_{Q\,Q_{2}}^{\mathcal{G}_{1}%
}\underline{\Gamma}_{\bar{Q}\,_{B}\{\bar{Q}_{1}Q\}}^{\mathcal{Q}_{2}}%
=\frac{g^{3}}{2k^{-}}\left[  -\bar{u}_{Q_{1}}\frac{t^{a}t^{\mathcal{G}_{1}%
}\tilde{F}_{4}^{-\mu}(k_{2}^{\prime\prime},k_{1})}{d(k_{2}^{\prime\prime
},k_{1})-m^{2}}\upsilon_{\bar{Q}_{2}}\otimes t^{a}\gamma_{\bot}^{\mu}u_{Q_{B}%
}\right.  \nonumber\\
&  +\left.  \frac{t^{a}t^{\mathcal{G}_{1}}\gamma_{\bot}^{\mu}}{x_{2}}%
\upsilon_{\bar{Q}_{2}}\otimes\bar{u}_{Q_{1}}\frac{t^{a}F_{3}^{-\mu}%
(p_{B}-k_{1},k_{1})}{d(p_{B}-k_{1},k_{1})-x_{2}^{2}m^{2}}u_{Q_{B}}\right]  \,.
\end{align}

Now one can clearly see that (\ref{bootstrap_QbarQ_frag}) is an identity.

\subsection{Two--gluon jet production in central region}

Here and in the following subsections we denote the momentum of the incoming
Reggeon as $q_{2}$ and  the momenta of the outgoing Reggeons $\mathcal{R}_{1}$
and $\mathcal{R}_{2}$ (the corresponding "bra"--vector $\langle\mathcal{R}%
_{1}\mathcal{R}_{2}|$) as $\,r_{1}$ and $r_{2}$ correspondingly, $r_{1}%
+r_{2}=q_{1}$. The momenta of the emitted particles (here they are gluons
$G_{1}$ and $G_{2}$) are $k_{1}$ and $k_{2}$ ($k_{1}+k_{2}=k$). The bootstrap
condition for $\gamma_{\mathcal{Q}_{1}\mathcal{Q}_{2}}^{\{G_{1}G_{2}\}}$ reads
as follows%
\begin{align}
&  \langle\mathcal{Q}_{1}\mathcal{G}_{2}|\,\widehat{\mathcal{\{}G_{1}%
G_{2}\mathcal{\}}}\mathcal{\,}|\mathcal{Q}_{\omega}(q_{2\bot})\rangle
\,g\,(m-\hat{q}_{2\bot})+\langle\mathcal{Q}_{1}\mathcal{G}_{2}|\{\bar{G}%
_{1}\bar{G}_{2}\}\mathcal{Q}_{2}\rangle\nonumber\\
&  =-\delta(q_{1\bot}+k_{\bot}-q_{2\bot})\,t^{\mathcal{G}_{2}}\mathcal{\,}%
g\,\gamma_{\mathcal{Q}_{1}\mathcal{Q}_{2}}^{\{G_{1}G_{2}\}}%
,\label{bootstrapGG}%
\end{align}
where $\widehat{\mathcal{\{}G_{1}G_{2}\mathcal{\}}}\mathcal{\,}$\ is the
operator of two gluon production with the matrix element
\begin{align*}
&  \langle\mathcal{Q}_{1}\mathcal{G}_{2}|\,\widehat{\mathcal{\{}G_{1}%
G_{2}\mathcal{\}}}|\mathcal{Q}_{1}^{\prime}\mathcal{G}_{2}^{\prime}%
\rangle=\delta(q_{1\bot}+k_{\bot}-q_{2\bot})\\
&  \times\left(  \,\gamma_{\mathcal{Q}_{1}\mathcal{Q}_{1}^{\prime}}%
^{\{G_{1}G_{2}\}}\delta_{\mathcal{G}_{2}\mathcal{G}_{2}^{\prime}}%
\delta(r_{2\bot}-r_{2\bot}^{\prime})r_{2\bot}^{2}\right.
\end{align*}%
\begin{align}
&  +\gamma_{\mathcal{G}_{2}\mathcal{G}_{2}^{\prime}}^{\{G_{1}G_{2}\}}%
\delta_{\mathcal{Q}_{1}\mathcal{Q}_{1}^{\prime}}\delta(r_{1\bot}-r_{1\bot
}^{\prime})(m-\hat{r}_{1\bot})+\gamma_{\mathcal{G}_{2}\mathcal{G}_{2}^{\prime
}}^{G_{2}}\gamma_{\mathcal{Q}_{1}\mathcal{Q}_{1}^{\prime}}^{G_{1}}%
\delta\left(  r_{2\bot}+k_{2\bot}-r_{2\bot}^{\prime}\right)
\label{OperatorGGProduction}\\
&  +\left.  \gamma_{\mathcal{G}_{2}\mathcal{G}_{2}^{\prime}}^{G_{1}}%
\gamma_{\mathcal{Q}_{1}\mathcal{Q}_{1}^{\prime}}^{G_{2}}\delta\left(
r_{2\bot}+k_{1\bot}-r_{2\bot}^{\prime}\right)  \right)  \,,\nonumber
\end{align}
and%
\begin{align}
&  \langle\mathcal{Q}_{1}\mathcal{G}_{2}|\{\bar{G}_{1}\bar{G}_{2}%
\}\mathcal{Q}_{2}\rangle=\delta(q_{1\bot}+k_{\bot}-q_{2\bot})\nonumber\\
&  \times\left[  \frac{1}{4k^{-}}\left(  \sum_{G}\Gamma_{\{G_{1}G_{2}%
\}G}^{\mathcal{G}_{2}}\,\gamma_{\mathcal{Q}_{1}\mathcal{Q}_{2}}^{G}+\sum
_{\bar{Q}}\underline{\Gamma}_{\bar{Q}\{\bar{G}_{1}\bar{G}_{2}\}}%
^{\mathcal{Q}_{1}}\,\gamma_{\bar{Q}}^{\mathcal{G}_{2}\mathcal{Q}_{2}}\right)
\right.  \nonumber\\
&  +\left.  \frac{1}{2k_{1}^{-}}\left(  \sum_{G}\Gamma_{G_{1}G}^{\mathcal{G}%
_{2}}\,\gamma_{\mathcal{Q}_{1}\mathcal{Q}_{2}}^{\{GG_{2}\}}+(-1)\sum_{\bar{Q}%
}\underline{\Gamma}_{\bar{Q}\,\bar{G}_{1}}^{\mathcal{Q}_{1}}\,\gamma
_{\{\bar{Q}G_{2}\}}^{\mathcal{G}_{2}\mathcal{Q}_{2}}\right)
+(1\leftrightarrow2)\right]  \,.\label{IFGGprodcentre}%
\end{align}
The term $\gamma_{\mathcal{G}_{2}\mathcal{G}_{2}^{\prime}}^{G_{2}}%
\gamma_{\mathcal{Q}_{1}\mathcal{Q}_{1}^{\prime}}^{G_{1}}\delta\left(
r_{2\bot}+k_{2\bot}-r_{2\bot}^{\prime}\right)  $ in bootstrap condition
(\ref{OperatorGGProduction}) is transformed via (\ref{id1}) as%
\begin{align}
&  \gamma_{\mathcal{G}_{2}\mathcal{G}_{2}^{\prime}}^{G_{2}}\gamma
_{\mathcal{Q}_{1}\mathcal{Q}_{1}^{\prime}}^{G_{1}}\delta\left(  r_{2\bot
}+k_{2\bot}-r_{2\bot}^{\prime}\right)  =\delta(q_{1\bot}+k_{\bot}-q_{2\bot
})\,g^{3}t^{G_{1}}\left[  t_{2}^{\mathcal{G}_{2}}t^{G_{2}}\right]  \nonumber\\
&  \times\left\{  e_{2\bot}^{\mu}\left[  \frac{{}}{{}}K_{2}^{\mu}\left(
k_{2}^{\prime},k_{1}^{\prime}\right)  +K_{1}^{\mu}(k_{2}^{\prime}%
,k_{1}^{\prime})-K_{2}^{\mu}\left(  k_{2},k_{1}^{\prime}\right)  -K_{1}^{\mu
}(k_{2},k_{1}^{\prime})\right]  \right.  \nonumber\\
&  \left.  +e_{1\bot}^{\mu}\left[  \frac{{}}{{}}K_{3}^{\mu}(k_{2}%
,k_{1})+\tilde{K}_{3}^{\mu}(k_{2},k_{1})-K_{3}^{\mu}(k_{2}^{\prime}%
,k_{1})-\tilde{K}_{3}^{\mu}(k_{2}^{\prime},k_{1})\right]  \right\}  (\hat
{q}_{2\bot}-m),
\end{align}
The similar procedure helps us to rewrite the contribution of $\gamma
_{\mathcal{G}_{2}\mathcal{G}_{2}^{\prime}}^{G_{1}}\gamma_{\mathcal{Q}%
_{1}\mathcal{Q}_{1}^{\prime}}^{G_{2}}\,.$ Performing cancellations (which are
trivial due to our effective vertex presentation) inside formulas
(\ref{OperatorGGProduction}) and (\ref{IFGGprodcentre}) we get%
\begin{align*}
&  \langle\mathcal{Q}_{1}\mathcal{G}_{2}|\,\,\widehat{\mathcal{\{}G_{1}%
G_{2}\mathcal{\}}}\mathcal{\,}|\mathcal{Q}_{\omega}(q_{2\bot})\rangle
\,g\,(m-\hat{q}_{2\bot})\\
&  =g^{3}\delta(q_{1\bot}+k_{\bot}-q_{2})\left\{  \left[  \left[  t^{G_{2}%
}t^{G_{1}}\right]  t^{\mathcal{G}_{2}}\right]  \frac{(q_{2}-r_{1})_{\bot}%
^{\mu}V_{1}^{\mu}(k_{2},k_{1})}{d(k_{2},k_{1})(q_{2}-r_{1})_{\bot}^{2}%
}\right.  \\
&  +2\left[  \left[  t^{\mathcal{G}_{2}}t^{G_{1}}\right]  t^{G_{2}}\right]
\frac{e_{1\bot}^{\mu}V_{2}^{\mu}(k_{2},k_{1}^{\prime\prime})}{D(k_{2}%
,k_{1}^{\prime\prime})(q_{2}-r_{1})_{\bot}^{2}}%
\end{align*}
\vspace{-0.9cm}
\begin{align*}
&  -\frac{t^{\mathcal{G}_{2}}\left[  t^{G_{1}}t^{G_{2}}\right]  }{(k_{1}%
+k_{2})_{\bot}^{2}}\left(  \frac{(k_{1}+k_{2})_{\bot}^{\mu}V_{1}^{\mu}%
(k_{2},k_{1})}{d(k_{2},k_{1})}+\frac{e_{1\bot}^{\mu}V_{2}^{\mu}(k_{2},k_{1}%
)}{D(k_{2},k_{1})}\right)  \\
&  +\left[  t^{G_{1}}t^{G_{2}}\right]  t^{\mathcal{G}_{2}}\frac{\gamma^{\mu
}V_{1}^{\mu}(k_{2},k_{1})}{2d(k_{2},k_{1})}\frac{1}{\hat{q}_{2\bot}-\hat
{r}_{2\bot}-m}%
\end{align*}
\vspace{-0.9cm}
\begin{align}
&  +t^{\mathcal{G}_{2}}t^{G_{2}}t^{G_{1}}e_{1\bot}^{\mu}K_{3}^{\mu}%
(k_{2},k_{1})-t^{G_{1}}t^{\mathcal{G}_{2}}t^{G_{2}}e_{2\bot}^{\mu}K_{2}^{\mu
}(k_{2},k_{1}^{\prime})\nonumber\\
&  +\left[  t^{G_{2}}t^{\mathcal{G}_{2}}\right]  t^{G_{1}}e_{1\bot}^{\mu}%
K_{3}^{\mu}(k_{2}^{\prime},k_{1})-t^{G_{2}}\left[  t^{\mathcal{G}_{2}}%
t^{G_{1}}\right]  e_{1\bot}^{\mu}K_{3}^{\mu}(k_{2},k_{1}^{\prime\prime
})\nonumber\\
&  +t^{G_{1}}\left[  t^{\mathcal{G}_{2}}t^{G_{2}}\right]  e_{2\bot}^{\mu
}\left(  \frac{{}}{{}}K_{2}^{\mu}\left(  k_{2}^{\prime},k_{1}^{\prime}\right)
+K_{1}^{\mu}(k_{2}^{\prime},k_{1}^{\prime})-K_{1}^{\mu}(k_{2},k_{1}^{\prime
})\right)  \nonumber\\
&  -\left.  t^{G_{1}}t^{G_{2}}t^{\mathcal{G}_{2}}\frac{e_{2\bot}^{\mu}%
F_{1}^{\mu}\left(  k_{2},k_{1}^{\prime}\right)  }{D\left(  k_{2},k_{1}%
^{\prime}\right)  -x_{2}m^{2}}\frac{1}{\hat{q}_{2\bot}-\hat{r}_{2\bot}%
-m}\right\}  (\hat{q}_{2\bot}-m)+(1\leftrightarrow2)\label{GGcentreRezult}%
\end{align}
and%
\begin{align*}
&  \langle\mathcal{Q}_{1}\mathcal{G}_{2}|\{\bar{G}_{1}\bar{G}_{2}%
\}\mathcal{Q}_{2}\rangle\\
&  =g^{3}\delta(q_{1\bot}+k_{\bot}-q_{2\bot})\left\{  t^{\mathcal{G}_{2}%
}\left[  t^{G_{1}}t^{G_{2}}\right]  \frac{\gamma^{\mu}V_{1}^{\mu}(k_{2}%
,k_{1})}{2d(k_{2},k_{1})}\right.  \\
&  -t^{\mathcal{G}_{2}}t^{G_{1}}t^{G_{2}}e_{2\bot}^{\mu}\left(  \frac
{F_{1}^{\mu}\left(  k_{2},q_{2}-k_{2}\right)  }{D\left(  k_{2},q_{2}%
-k_{2}\right)  -x_{2}m^{2}}+K_{2}^{\mu}\left(  k_{2},q_{2}-k_{2}\right)
(\hat{q}_{2\bot}-m)\right)
\end{align*}
\vspace{-0.8cm}
\begin{align*}
&  +\left(  \left[  \left[  t^{G_{1}}t^{G_{2}}\right]  t^{\mathcal{G}_{2}%
}\right]  \frac{(q_{2}-r_{1})_{\bot}^{\mu}V_{1}^{\mu}(k_{2},k_{1})}%
{d(k_{2},k_{1})(q_{2}-r_{1})_{\bot}^{2}}-\left[  t^{G_{1}}t^{G_{2}}\right]
t^{\mathcal{G}_{2}}\frac{\gamma^{\mu}V_{1}^{\mu}(k_{2},k_{1})}{2d(k_{2}%
,k_{1})}\frac{1}{\hat{q}_{2\bot}-\hat{r}_{2\bot}-m}\right.  \\
&  -2\left[  t^{G_{1}}\left[  t^{\mathcal{G}_{2}}t^{G_{2}}\right]  \right]
\frac{e_{1\bot}^{\mu}V_{2}^{\mu}(k_{2}^{\prime},k_{1})}{D(k_{2}^{\prime}%
,k_{1})(q_{2}-r_{1})_{\bot}^{2}}+t^{G_{1}}t^{G_{2}}t^{\mathcal{G}_{2}}%
\frac{e_{2\bot}^{\mu}F_{1}^{\mu}\left(  k_{2},k_{1}^{\prime}\right)
}{D\left(  k_{2},k_{1}^{\prime}\right)  -x_{2}m^{2}}\frac{1}{\hat{q}_{2\bot
}-\hat{r}_{2\bot}-m}%
\end{align*}
\vspace*{-1cm}
\begin{align}
&  +t^{G_{1}}t^{\mathcal{G}_{2}}t^{G_{2}}e_{2\bot}^{\mu}K_{2}^{\mu}%
(k_{2},k_{1}^{\prime})+t^{G_{1}}\left[  t^{\mathcal{G}_{2}}t^{G_{2}}\right]
e_{1\bot}^{\mu}\tilde{K}_{3}^{\mu}(k_{2}^{\prime},k_{1})\nonumber\\
&  -t^{G_{1}}\left[  t^{\mathcal{G}_{2}}t^{G_{2}}\right]  e_{2\bot}^{\mu
}\left(  \frac{{}}{{}}K_{2}^{\mu}\left(  k_{2}^{\prime},k_{1}^{\prime}\right)
+K_{1}^{\mu}(k_{2}^{\prime},k_{1}^{\prime})-K_{1}^{\mu}(k_{2},k_{1}^{\prime
})\right)  \nonumber\\
&  +\left.  \left.  \left[  t^{\mathcal{G}_{2}}t^{G_{2}}\right]  t^{G_{1}%
}e_{1\bot}^{\mu}K_{3}^{\mu}(k_{2}^{\prime},k_{1})\frac{{}}{{}}\right)
(\hat{q}_{2\bot}-m)\right\}  \,+(1\leftrightarrow
2)\,\,.\label{IFGGpCentRezult}%
\end{align}
One can easily check that the sum of (\ref{GGcentreRezult}) and
(\ref{IFGGpCentRezult}) gives the r.h.s. of (\ref{bootstrapGG}) fulfilling
bootstrap condition (\ref{bootstrap production ket}) for this case.

\subsection{Quark--antiquark jet production in central region}

We denote the momenta of the emitted quark and antiquark as
$k_{1}$ and $k_{2}$ respectively.
The bootstrap condition for $\gamma_{\mathcal{Q}_{1}\mathcal{Q}_{2}}%
^{\{Q_{1}\bar{Q}_{2}\}}$ has the form:%
\begin{align}
&  \langle\mathcal{Q}_{1}\mathcal{G}_{2}|\,\widehat{\{Q_{1}\bar{Q}_{2}%
\}}\,|\mathcal{Q}_{\omega}(q_{2\bot})\rangle\,g\,(m-\hat{q}_{2\bot}%
)+\langle\mathcal{Q}_{1}\mathcal{G}_{2}|\{\bar{Q}_{1}Q_{2}\}\mathcal{Q}%
_{2}\rangle\nonumber\\
&  =\langle\mathcal{Q}_{1}\mathcal{G}_{2}|\mathcal{Q}_{\omega}(q_{1\bot
})\rangle\mathcal{\,}g\,\gamma_{\mathcal{Q}_{1}\mathcal{Q}_{2}}^{\{Q_{1}%
\bar{Q}_{2}\}},\label{bootstrap_antiQQcentr}%
\end{align}
where $\widehat{\{Q_{1}\bar{Q}_{2}\}}\mathcal{\,}$ is the operator of
quark-antiquark production with the matrix elements%
\begin{align}
&  \langle\mathcal{Q}_{1}\mathcal{G}_{2}|\widehat{\{Q_{1}\bar{Q}_{2}%
\}}\mathcal{\,}|\mathcal{Q}_{1}^{\prime}\mathcal{G}_{2}^{\prime}\rangle
=\delta(q_{1\bot}+k_{\bot}-q_{2\bot})\nonumber\\
&  \times\left[  \gamma_{\mathcal{Q}_{1}\mathcal{Q}_{1}^{\prime}}^{\{Q_{1}%
\bar{Q}_{2}\}}\delta_{\mathcal{G}_{2}\mathcal{G}_{2}^{\prime}}\delta(r_{2\bot
}-r_{2\bot}^{\prime})r_{2\bot}^{2}\right.  \nonumber\\
&  +\left.  \gamma_{\mathcal{G}_{2}\mathcal{G}_{2}^{\prime}}^{\{Q_{1}\bar
{Q}_{2}\}}\delta_{\mathcal{Q}_{1}\mathcal{Q}_{1}^{\prime}}\delta(r_{1\bot
}-r_{1\bot}^{\prime})(m-\hat{r}_{1\bot})\right]  ,
\end{align}%
\begin{align}
&  \langle\mathcal{Q}_{1}\mathcal{G}_{2}|\,\widehat{\{Q_{1}\bar{Q}_{2}%
\}}|\mathcal{G}_{1}^{\prime}\mathcal{Q}_{2}^{\prime}\rangle\nonumber\\
&  =\delta(q_{1\bot}+k_{\bot}-q_{2})\,\,\gamma_{\mathcal{Q}_{1}\mathcal{G}%
_{1}^{\prime}}^{\bar{Q}_{2}}\gamma_{\mathcal{G}_{2}\mathcal{Q}_{2}^{\prime}%
}^{Q_{1}}\delta\left(  r_{2\bot}+k_{1\bot}-r_{2\bot}^{\prime}\right)
\end{align}
and%
\begin{align}
&  \langle\mathcal{Q}_{1}\mathcal{G}_{2}|\{\bar{Q}_{1}Q_{2}\}\mathcal{Q}%
_{2}\rangle=\delta(q_{1\bot}+k_{\bot}-q_{2\bot})\nonumber\\
&  \times\left[  \frac{1}{2k^{-}}\left(  \sum_{G}\Gamma_{\{Q_{1}\bar{Q}%
_{2}\}G}^{\mathcal{G}_{2}}\,\gamma_{\mathcal{Q}_{1}\mathcal{Q}_{2}}^{G}%
+\sum_{\bar{Q}}\underline{\Gamma}_{\bar{Q}\{\bar{Q}_{1}Q_{2}\}}^{\mathcal{Q}%
_{1}}\,\gamma_{\bar{Q}}^{\mathcal{G}_{2}\mathcal{Q}_{2}}\right)  \right.
\nonumber\\
&  +\left.  \frac{1}{2k_{1}^{-}}\sum_{Q}\Gamma_{Q_{1}Q}^{\mathcal{G}_{2}%
}\,\gamma_{\mathcal{Q}_{1}\mathcal{Q}_{2}}^{\{Q\bar{Q}_{2}\}}+\frac{1}%
{2k_{2}^{-}}\left(  \sum_{\bar{Q}}\Gamma_{\bar{Q}_{2}\bar{Q}}^{\mathcal{G}%
_{2}}\,\gamma_{\mathcal{Q}_{1}\mathcal{Q}_{2}}^{\{Q_{1}\bar{Q}\}}+\sum
_{G}\underline{\Gamma}_{G\,Q_{2}}^{\mathcal{Q}_{1}}\,\gamma_{\{\bar{Q}_{1}%
G\}}^{\mathcal{G}_{2}\mathcal{Q}_{2}}\right)  \right]  \,.
\end{align}
With the help of (\ref{id1}) the contribution of $\,\gamma_{\mathcal{Q}%
_{1}\mathcal{G}_{1}^{\prime}}^{\bar{Q}_{2}}\gamma_{\mathcal{G}_{2}%
\mathcal{Q}_{2}^{\prime}}^{Q_{1}}$ into $\langle\mathcal{Q}_{1}\mathcal{G}%
_{2}|\,\widehat{\{Q_{1}\bar{Q}_{2}\}}\mathcal{\,}|\mathcal{Q}_{\omega
}(q_{2\bot})\rangle\,g\,(m-\hat{q}_{2\bot})$ yields%
\begin{align}
&  \delta(q_{1\bot}+k_{\bot}-q_{2\bot})\,g^{3}t^{\mathcal{G}_{1}^{\prime}%
}\frac{\gamma_{\bot}^{\mu}}{2k_{2}^{-}}\upsilon_{\bar{Q}_{2}}%
\nonumber\label{bootstrap_antiQQcent}\\
&  \otimes\bar{u}_{Q_{1}}t^{\mathcal{G}_{2}}t^{\mathcal{G}_{1}^{\prime}%
}\left(  \frac{{}}{{}}K_{1}^{-\mu}\left(  k_{2}^{\prime\prime},k_{1}\right)
+K_{2}^{-\mu}(k_{2}^{\prime\prime},k_{1})\right.  \nonumber\\
&  \left.  -\frac{{}}{{}}K_{1}^{-\mu}\left(  k_{2}^{\prime\prime}%
,k_{1}^{\prime\prime}\right)  -K_{2}^{-\mu}(k_{2}^{\prime\prime},k_{1}%
^{\prime\prime})\right)  (\hat{q}_{2\bot}-m)\,.
\end{align}
We can present the result for $\langle\mathcal{Q}_{1}\mathcal{G}_{2}|\{\bar
{Q}_{1}Q_{2}\}\mathcal{Q}_{2}\rangle$ in the following form:%
\begin{align*}
&  \langle\mathcal{Q}_{1}\mathcal{G}_{2}|\{\bar{Q}_{1}Q_{2}\}\mathcal{Q}%
_{2}\rangle=\delta(q_{1\bot}+k_{\bot}-q_{2\bot})\frac{g^{3}}{k^{-}}\left[
\,t^{a}\frac{\gamma_{\bot}^{\mu}}{2x_{2}}\upsilon_{\bar{Q}_{2}}\right.  \\
&  \otimes\bar{u}_{Q_{1}}\left\{  t^{\mathcal{G}_{2}}t^{a}\left(  \frac
{F_{1}^{-\mu}\left(  q_{2}-k_{1},k_{1}\right)  }{D\left(  q_{2}-k_{1}%
,k_{1}\right)  -x_{2}m^{2}}-K_{1}^{-\mu}\left(  q_{2}-k_{1},k_{1}\right)
(\hat{q}_{2\bot}-m)\right)  \right.  \\
&  -t^{\mathcal{G}_{2}}t^{a}\left(  \frac{{}}{{}}K_{1}^{-\mu}\left(
k_{2}^{\prime\prime},k_{1}\right)  +K_{2}^{-\mu}(k_{2}^{\prime\prime}%
,k_{1})-K_{2}^{-\mu}(k_{2}^{\prime\prime},k_{1}^{\prime\prime})-K_{1}^{-\mu
}(k_{2}^{\prime\prime},k_{1}^{\prime\prime})\right)
\end{align*}
\vspace{-0.9cm}
\begin{align*}
&  (\hat{q}_{2\bot}-m)\\
&  \left.  -t^{a}t^{\mathcal{G}_{2}}\left(  \frac{F_{1}^{-\mu}\left(
k_{2}^{\prime\prime},k_{1}\right)  }{D\left(  k_{2}^{\prime\prime}%
,k_{1}\right)  -x_{2}m^{2}}\frac{1}{\hat{q}_{2\bot}-\hat{r}_{2\bot}-m}%
-K_{1}^{-\mu}\left(  k_{2}^{\prime\prime},k_{1}\right)  (\hat{q}_{2\bot
}-m)\right)  \right\}  \\
&  +(\hat{q}_{2\bot}-m)\,t^{a}%
\end{align*}
\vspace{-0.9cm}
\begin{align}
&  \otimes\frac{\bar{u}_{Q_{1}}\hat{n}_{1}}{\left(  q_{2}-r_{1}\right)
_{\bot}^{2}}\left(  \left[  t^{a}t^{\mathcal{G}_{2}}\right]  +\frac
{t^{a}t^{\mathcal{G}_{2}}F_{5}\left(  k_{2}^{\prime},k_{1}\right)  }{D\left(
k_{2}^{\prime},k_{1}\right)  -m^{2}}-\frac{t^{\mathcal{G}_{2}}t^{a}%
F_{5}\left(  k_{2},k_{1}^{\prime\prime}\right)  }{D\left(  k_{2},k_{1}%
^{\prime\prime}\right)  -m^{2}}\right)  \upsilon_{\bar{Q}_{2}}\nonumber\\
&  +\,\left\{  \frac{\gamma_{\bot}^{\mu}}{2}\left(  t^{\mathcal{G}_{2}}%
t^{a}-\frac{t^{a}t^{\mathcal{G}_{2}}}{\hat{q}_{2\bot}-\hat{r}_{2\bot}-m}%
(\hat{q}_{2\bot}-m)\right)  -\left[  t^{\mathcal{G}_{2}}t^{a}\right]
\frac{(\hat{q}_{2\bot}-m)\left(  q_{2}-r_{1}\right)  _{\bot}^{\mu}}{\left(
q_{2}-r_{1}\right)  _{\bot}^{2}}\right\}  \nonumber\\
&  \otimes\left.  \bar{u}_{Q_{1}}\frac{t^{a}\tilde{F}_{4}^{-\mu}\left(
k_{2},k_{1}\right)  }{d\left(  k_{2},k_{1}\right)  -m^{2}}\upsilon_{\bar
{Q}_{2}}\right]  .
\end{align}
The contribution of $\langle\mathcal{Q}_{1}\mathcal{G}_{2}|\,\widehat
{\{Q_{1}\bar{Q}_{2}\}}\mathcal{\,}|\mathcal{Q}_{\omega}(q_{2\bot}%
)\rangle\,g\,(m-\hat{q}_{2\bot})$ into the bootstrap relation reads as
follows:%
\begin{align*}
&  \langle\mathcal{Q}_{1}\mathcal{G}_{2}|\,\widehat{\{Q_{1}\bar{Q}_{2}%
\}}\mathcal{\,}|\mathcal{Q}_{\omega}(q_{2\bot})\rangle\,g\,(m-\hat{q}_{2\bot
})\\
&  =\delta(q_{1\bot}+k_{\bot}-q_{2\bot})\frac{g^{3}}{k^{-}}\left[
\,t^{a}\frac{\gamma_{\bot}^{\mu}}{2x_{2}}\upsilon_{\bar{Q}_{2}}\right.  \\
&  \otimes\bar{u}_{Q_{1}}\left\{  t^{a}t^{\mathcal{G}_{2}}\left(  \frac
{F_{1}^{-\mu}\left(  k_{2}^{\prime\prime},k_{1}\right)  }{D\left(
k_{2}^{\prime\prime},k_{1}\right)  -x_{2}m^{2}}\frac{1}{\hat{q}_{2\bot}%
-\hat{r}_{2\bot}-m}-K_{1}^{-\mu}\left(  k_{2}^{\prime\prime},k_{1}\right)
(\hat{q}_{2\bot}-m)\right)  \right.
\end{align*}
\vspace{-0.9cm}
\begin{align*}
&  +\left.  t^{\mathcal{G}_{2}}t^{a}\left(  \frac{{}}{{}}K_{1}^{-\mu}\left(
k_{2}^{\prime\prime},k_{1}\right)  +K_{2}^{-\mu}(k_{2}^{\prime\prime}%
,k_{1})-K_{1}^{-\mu}\left(  k_{2}^{\prime\prime},k_{1}^{\prime\prime}\right)
-K_{2}^{-\mu}(k_{2}^{\prime\prime},k_{1}^{\prime\prime})\right)  \right\}  \\
&  (\hat{q}_{2\bot}-m)-(\hat{q}_{2\bot}-m)\,t^{a}\\
&  \otimes\frac{\bar{u}_{Q_{1}}\hat{n}_{1}}{\left(  q_{2}-r_{1}\right)
_{\bot}^{2}}\left(  \left[  t^{a}t^{\mathcal{G}_{2}}\right]  +\frac
{t^{a}t^{\mathcal{G}_{2}}F_{5}\left(  k_{2}^{\prime},k_{1}\right)  }{D\left(
k_{2}^{\prime},k_{1}\right)  -m^{2}}-\frac{t^{\mathcal{G}_{2}}t^{a}%
F_{5}\left(  k_{2},k_{1}^{\prime\prime}\right)  }{D\left(  k_{2},k_{1}%
^{\prime\prime}\right)  -m^{2}}\right)  \upsilon_{\bar{Q}_{2}}%
\end{align*}
\vspace{-0.9cm}
\begin{align}
&  +\left\{  \frac{\gamma_{\bot}^{\mu}}{2}\frac{t^{a}t^{\mathcal{G}_{2}}}%
{\hat{q}_{2\bot}-\hat{r}_{2\bot}-m}+\left[  t^{\mathcal{G}_{2}}t^{a}\right]
\frac{\left(  q_{2}-r_{1}\right)  _{\bot}^{\mu}}{\left(  q_{2}-r_{1}\right)
_{\bot}^{2}}\right\}  (\hat{q}_{2\bot}-m)\nonumber\\
&  \otimes\,\bar{u}_{Q_{1}}\frac{t^{a}\tilde{F}_{4}^{-\mu}\left(  k_{2}%
,k_{1}\right)  }{d\left(  k_{2},k_{1}\right)  -m^{2}}\upsilon_{\bar{Q}_{2}%
}+(\hat{q}_{2\bot}-m)\,t^{\mathcal{G}_{2}}t^{a}\nonumber\\
&  \otimes\left.  \frac{\bar{u}_{Q_{1}}t^{a}}{\left(  k_{1}+k_{2}\right)
_{\bot}^{2}}\left\{  \hat{n}_{1}+\frac{\hat{n}_{1}F_{5}\left(  k_{2}%
,k_{1}\right)  }{D\left(  k_{2},k_{1}\right)  -m^{2}}-\frac{\left(
k_{1}+k_{2}\right)  _{\bot}^{\mu}\tilde{F}_{4}^{-\mu}\left(  k_{2}%
,k_{1}\right)  }{d\left(  k_{2},k_{1}\right)  -m^{2}}\right\}  \upsilon
_{\bar{Q}_{2}}\right]  \,.
\end{align}
The sum of the two last expressions gives the r.h.s. of
(\ref{bootstrap_antiQQcentr}) fulfilling bootstrap condition for
$\gamma_{\mathcal{Q}_{1}\mathcal{Q}_{2}}^{\{Q_{1}\bar{Q}_{2}\}}$.

\subsection{Quark--gluon jet production in central region}

We denote the momenta of the emitted quark and gluon as $k_{1}$ and $k_{2}%
$. The bootstrap condition for $\gamma_{\mathcal{G}_{1}\mathcal{Q}_{2}}%
^{\{Q_{1}G_{2}\}}$ has the form:%
\begin{align}
&  \langle\mathcal{G}_{1}\mathcal{G}_{2}|\,\widehat{\{Q_{1}G_{2}%
\}}\mathcal{\,}|\mathcal{Q}_{\omega}(q_{2\bot})\rangle\,g\,(m-\hat{q}_{2\bot
})+\langle\mathcal{G}_{1}\mathcal{G}_{2}|\{\bar{Q}_{1}\bar{G}_{2}%
\}\mathcal{Q}_{2}\rangle\nonumber\\
&  =\langle\mathcal{G}_{1}\mathcal{G}_{2}|\mathcal{G}_{\omega}(q_{1\bot
})\rangle\mathcal{\,}g\,\gamma_{\mathcal{G}_{1}\mathcal{Q}_{2}}^{\{Q_{1}%
G_{2}\}},\label{bootstrapQG}%
\end{align}
where $\widehat{\{Q_{1}G_{2}\}}\mathcal{\,}$ is quark--gluon production
operator with the matrix elements%
\begin{align}
&  \langle\mathcal{G}_{1}\mathcal{G}_{2}|\,\widehat{\{Q_{1}G_{2}%
\}}\mathcal{\,}|\mathcal{Q}_{1}^{\prime}\mathcal{G}_{2}^{\prime}\rangle
=\delta(q_{1\bot}+k_{\bot}-q_{2\bot})\nonumber\\
&  \times\left[  \gamma_{\mathcal{G}_{1}\mathcal{Q}_{1}^{\prime}}%
^{\{Q_{1}G_{2}\}}\delta_{\mathcal{G}_{2}\mathcal{G}_{2}^{\prime}}%
\delta(r_{2\bot}-r_{2\bot}^{\prime})r_{2\bot}^{2}+\gamma_{\mathcal{G}%
_{2}\mathcal{G}_{2}^{\prime}}^{G_{2}}\gamma_{\mathcal{G}_{1}\mathcal{Q}%
_{1}^{\prime}}^{Q_{1}}\delta\left(  r_{2\bot}+k_{2\bot}-r_{2\bot}^{\prime
}\right)  \right]  ,
\end{align}%
\begin{align}
&  \langle\mathcal{G}_{1}\mathcal{G}_{2}|\,\widehat{\{Q_{1}G_{2}%
\}}\mathcal{\,}|\mathcal{G}_{1}^{\prime}\mathcal{Q}_{2}^{\prime}\rangle
=\delta(q_{1\bot}+k_{\bot}-q_{2\bot})\nonumber\\
&  \times\left[  \gamma_{\mathcal{G}_{2}\mathcal{Q}_{2}^{\prime}}%
^{\{Q_{1}G_{2}\}}\delta_{\mathcal{G}_{1}\mathcal{G}_{1}^{\prime}}%
\delta(r_{1\bot}-r_{1\bot}^{\prime})r_{1\bot}^{2}+\gamma_{\mathcal{G}%
_{1}\mathcal{G}_{1}^{\prime}}^{G_{2}}\gamma_{\mathcal{G}_{2}\mathcal{Q}%
_{2}^{\prime}}^{Q_{1}}\delta\left(  r_{2\bot}+k_{1\bot}-r_{2\bot}^{\prime
}\right)  \right]
\end{align}
and%
\begin{align*}
&  \langle\mathcal{G}_{1}\mathcal{G}_{2}|\{\bar{Q}_{1}\bar{G}_{2}%
\}\mathcal{Q}_{2}\rangle=\delta(q_{1\bot}+k_{\bot}-q_{2\bot})\\
&  \times\left[  \frac{1}{2k^{-}}\left(  \sum_{Q}\Gamma_{\{Q_{1}G_{2}%
\}Q}^{\mathcal{G}_{2}}\,\gamma_{\mathcal{G}_{1}\mathcal{Q}_{2}}^{Q}-\sum
_{\bar{Q}}\underline{\Gamma}_{\bar{Q}\{\bar{Q}_{1}\bar{G}_{2}\}}%
^{\mathcal{G}_{1}}\,\gamma_{\bar{Q}}^{\mathcal{G}_{2}\mathcal{Q}_{2}}\right)
\right.
\end{align*}%
\begin{align}
&  +\frac{1}{2k_{1}^{-}}\left(  \sum_{Q}\Gamma_{Q_{1}Q}^{\mathcal{G}_{2}%
}\,\gamma_{\mathcal{G}_{1}\mathcal{Q}_{2}}^{\{QG_{2}\}}-\sum_{\bar{Q}%
}\underline{\Gamma}_{\bar{Q}\,\bar{Q}_{1}}^{\mathcal{G}_{1}}\,\gamma
_{\{\bar{Q}\bar{G}_{2}\}}^{\mathcal{G}_{2}\mathcal{Q}_{2}}\right)  \nonumber\\
&  \left.  +\frac{1}{2k_{2}^{-}}\sum_{G}\left(  \Gamma_{G_{2}G}^{\mathcal{G}%
_{2}}\,\gamma_{\mathcal{G}_{1}\mathcal{Q}_{2}}^{\{Q_{1}G\}}-\underline{\Gamma
}_{G\,\bar{G}_{2}}^{\mathcal{G}_{1}}\,\gamma_{\{\bar{Q}G\}}^{\mathcal{G}%
_{2}\mathcal{Q}_{2}}\right)  \right]  .
\end{align}

The contribution of $\gamma_{\mathcal{G}_{1}\mathcal{G}_{1}^{\prime}}^{G_{2}%
}\gamma_{\mathcal{G}_{2}\mathcal{Q}_{2}^{\prime}}^{Q_{1}}$ into $\langle
\mathcal{G}_{1}\mathcal{G}_{2}|\,\widehat{\{Q_{1}G_{2}\}}\mathcal{\,}%
|\mathcal{Q}_{\omega}(q_{2\bot})\rangle\,g\,(m-\hat{q}_{2\bot})$ yields%
\begin{align}
&  \delta(q_{1\bot}+k_{\bot}-q_{2\bot})\,g^{3}t^{\mathcal{G}_{2}}\left[
t^{\mathcal{G}_{1}}t^{G_{2}}\right]  \nonumber\\
&  \times e_{2\bot}^{\mu}\bar{u}_{Q_{1}}\left(  \frac{{}}{{}}K_{2}^{\mu
}\left(  k_{2}^{\prime\prime},k_{1}^{\prime\prime}\right)  +K_{1}^{\mu}%
(k_{2}^{\prime\prime},k_{1}^{\prime\prime})-K_{2}^{\mu}\left(  k_{2}%
^{\prime\prime},k_{1}\right)  -K_{1}^{\mu}(k_{2}^{\prime\prime},k_{1})\right.
\nonumber\\
&  -\left.  \frac{{}}{{}}K_{2}^{\mu}\left(  k_{2},k_{1}^{\prime\prime}\right)
-K_{1}^{\mu}(k_{2},k_{1}^{\prime\prime})+K_{2}^{\mu}\left(  k_{2}%
,k_{1}\right)  +K_{1}^{\mu}(k_{2},k_{1})\right)  (\hat{q}_{2\bot
}-m)\label{gggq}%
\end{align}
and the contribution of $\gamma_{\mathcal{G}_{2}\mathcal{G}_{2}^{\prime}%
}^{G_{2}}\gamma_{\mathcal{G}_{1}\mathcal{Q}_{1}^{\prime}}^{Q_{1}}$ can be
obtained from (\ref{gggq}) by the substitution $r_{1}\leftrightarrow r_{2}$,
$\mathcal{G}_{1}\leftrightarrow\mathcal{G}_{2}$.

We can present the result for $\langle\mathcal{G}_{1}\mathcal{G}_{2}|\{\bar
{Q}_{1}\bar{G}_{2}\}\mathcal{Q}_{2}\rangle$ in the following form:%
\begin{align*}
&  \langle\mathcal{G}_{1}\mathcal{G}_{2}|\{\bar{Q}_{1}\bar{G}_{2}%
\}\mathcal{Q}_{2}\rangle\\
&  =\delta(q_{1\bot}+k_{\bot}-q_{2\bot})g^{3}e_{2\bot}^{\mu}\bar{u}_{Q_{1}%
}\left(  -t^{G_{2}}\left[  t^{\mathcal{G}_{2}}t^{\mathcal{G}_{1}}\right]
\frac{F_{3}^{\mu}\left(  k_{2},k_{1}\right)  }{d\left(  k_{2},k_{1}\right)
-x_{2}^{2}m^{2}}\right.  \\
&  +\left[  \left[  t^{\mathcal{G}_{2}}t^{\mathcal{G}_{1}}\right]  t^{G_{2}%
}\right]  \left(  \frac{F_{1}^{\mu}\left(  q_{2}-k_{1},k_{1}\right)
}{D\left(  q_{2}-k_{1},k_{1}\right)  -x_{2}m^{2}}-K_{1}^{\mu}\left(
q_{2}-k_{1},k_{1}\right)  (\hat{q}_{2\bot}-m)\right)
\end{align*}
\vspace{-0.9cm}
\begin{align*}
&  -\left[  t^{\mathcal{G}_{2}}t^{\mathcal{G}_{1}}\right]  t^{G_{2}}\left(
\frac{F_{1}^{\mu}\left(  k_{2},q_{2}-k_{2}\right)  }{D\left(  k_{2}%
,q_{2}-k_{2}\right)  -x_{2}m^{2}}+K_{2}^{\mu}\left(  k_{2},q_{2}-k_{2}\right)
(\hat{q}_{2\bot}-m)\right)  \\
&  +\left\{  \frac{F_{3}^{\mu}\left(  k_{2},k_{1}\right)  }{d\left(
k_{2},k_{1}\right)  -x_{2}^{2}m^{2}}\frac{t^{G_{2}}t^{\mathcal{G}_{2}%
}t^{\mathcal{G}_{1}}}{\hat{q}_{2\bot}-\hat{r}_{1\bot}-m}-\frac{F_{3}^{\mu
}\left(  k_{2},k_{1}\right)  }{d\left(  k_{2},k_{1}\right)  -x_{2}^{2}m^{2}%
}\frac{t^{G_{2}}t^{\mathcal{G}_{1}}t^{\mathcal{G}_{2}}}{\hat{q}_{2\bot}%
-\hat{r}_{2\bot}-m}\right.  \\
&  +t^{\mathcal{G}_{2}}\left[  t^{\mathcal{G}_{1}}t^{G_{2}}\right]  \left(
\frac{{}}{{}}K_{1}^{\mu}(k_{2},k_{1}^{\prime\prime})-K_{1}^{\mu}(k_{2}%
^{\prime\prime},k_{1}^{\prime\prime})\right.
\end{align*}
\vspace{-0.9cm}
\begin{align*}
&  +\left.  \frac{{}}{{}}K_{1}^{\mu}\left(  k_{2}^{\prime\prime},k_{1}\right)
+K_{2}^{\mu}(k_{2}^{\prime\prime},k_{1})-K_{2}^{\mu}(k_{2}^{\prime\prime
},k_{1}^{\prime\prime})\right)  \\
&  +\frac{F_{1}^{\mu}\left(  k_{2},k_{1}^{\prime\prime}\right)  }{D\left(
k_{2},k_{1}^{\prime\prime}\right)  -x_{2}m^{2}}\frac{t^{\mathcal{G}_{2}%
}t^{G_{2}}t^{\mathcal{G}_{1}}}{\hat{q}_{2\bot}-\hat{r}_{1\bot}-m}-\frac
{F_{1}^{\mu}\left(  k_{2},k_{1}^{\prime}\right)  }{D\left(  k_{2}%
,k_{1}^{\prime}\right)  -x_{2}m^{2}}\frac{t^{\mathcal{G}_{1}}t^{G_{2}%
}t^{\mathcal{G}_{2}}}{\hat{q}_{2\bot}-\hat{r}_{2\bot}-m}\\
&  +\left[  t^{G_{2}}t^{\mathcal{G}_{2}}\right]  t^{\mathcal{G}_{1}}\left(
\frac{F_{1}^{\mu}\left(  k_{2}^{\prime},k_{1}\right)  }{D\left(  k_{2}%
^{\prime},k_{1}\right)  -x_{2}m^{2}}\frac{1}{\hat{q}_{2\bot}-\hat{r}_{1\bot
}-m}-K_{1}^{\mu}\left(  k_{2}^{\prime},k_{1}\right)  \right)
\end{align*}
\vspace{-0.9cm}
\begin{align}
&  -\left[  t^{G_{2}}t^{\mathcal{G}_{1}}\right]  t^{\mathcal{G}_{2}}\left(
\frac{F_{1}^{\mu}\left(  k_{2}^{\prime\prime},k_{1}\right)  }{D\left(
k_{2}^{\prime\prime},k_{1}\right)  -x_{2}m^{2}}\frac{1}{\hat{q}_{2\bot}%
-\hat{r}_{2\bot}-m}-K_{1}^{\mu}\left(  k_{2}^{\prime\prime},k_{1}\right)
\right)  \nonumber\\
&  +t^{\mathcal{G}_{2}}t^{\mathcal{G}_{1}}t^{G_{2}}K_{2}^{\mu}(k_{2}%
,k_{1}^{\prime\prime})-t^{\mathcal{G}_{1}}t^{\mathcal{G}_{2}}t^{G_{2}}%
K_{2}^{\mu}(k_{2},k_{1}^{\prime})\nonumber\\
&  -t^{\mathcal{G}_{1}}\left[  t^{\mathcal{G}_{2}}t^{G_{2}}\right]  \left(
\frac{{}}{{}}K_{1}^{\mu}(k_{2},k_{1}^{\prime})-K_{1}^{\mu}(k_{2}^{\prime
},k_{1}^{\prime})+K_{1}^{\mu}\left(  k_{2}^{\prime},k_{1}\right)  \right.
\nonumber\\
&  +\left.  \left.  \left.  \frac{{}}{{}}K_{2}^{\mu}(k_{2}^{\prime}%
,k_{1})-K_{2}^{\mu}(k_{2}^{\prime},k_{1}^{\prime})\right)  \right\}  (\hat
{q}_{2\bot}-m)\frac{{}}{{}}\right)  \,.
\end{align}
The contribution of $\langle\mathcal{G}_{1}\mathcal{G}_{2}|\,\widehat
{\{Q_{1}G_{2}\}}\mathcal{\,}|\mathcal{Q}_{\omega}(q_{2\bot})\rangle
\,g\,(m-\hat{q}_{2\bot})$ into the bootstrap relation reads as follows:%
\begin{align*}
&  \langle\mathcal{G}_{1}\mathcal{G}_{2}|\,\widehat{\{Q_{1}G_{2}%
\}}\mathcal{\,}|\mathcal{Q}_{\omega}(q_{2\bot})\rangle\,g\,(m-\hat{q}_{2\bot
})=\delta(q_{1\bot}+k_{\bot}-q_{2\bot})g^{3}e_{2\bot}^{\mu}\bar{u}_{Q_{1}}\\
&  \times\left(  \frac{{}}{{}}\left[  t^{\mathcal{G}_{2}}t^{\mathcal{G}_{1}%
}\right]  t^{G_{2}}\left(  K_{1}^{\mu}\left(  k_{2},k_{1}\right)  +K_{2}^{\mu
}(k_{2},k_{1})\right)  \right.  \\
&  -t^{\mathcal{G}_{2}}\left[  t^{\mathcal{G}_{1}}t^{G_{2}}\right]  \left(
\frac{{}}{{}}K_{1}^{\mu}(k_{2},k_{1}^{\prime\prime})-K_{1}^{\mu}(k_{2}%
^{\prime\prime},k_{1}^{\prime\prime})+K_{1}^{\mu}\left(  k_{2}^{\prime\prime
},k_{1}\right)  +K_{2}^{\mu}(k_{2}^{\prime\prime},k_{1})-K_{2}^{\mu}%
(k_{2}^{\prime\prime},k_{1}^{\prime\prime})\right)
\end{align*}
\vspace{-0.9cm}
\begin{align*}
&  +t^{\mathcal{G}_{1}}\left[  t^{\mathcal{G}_{2}}t^{G_{2}}\right]  \left(
\frac{{}}{{}}K_{1}^{\mu}(k_{2},k_{1}^{\prime})-K_{1}^{\mu}(k_{2}^{\prime
},k_{1}^{\prime})+K_{1}^{\mu}\left(  k_{2}^{\prime},k_{1}\right)  +K_{2}^{\mu
}(k_{2}^{\prime},k_{1})-K_{2}^{\mu}(k_{2}^{\prime},k_{1}^{\prime})\right)  \\
&  +t^{G_{2}}\left[  t^{\mathcal{G}_{2}}t^{\mathcal{G}_{1}}\right]  \left\{
\left(  \frac{F_{3}^{\mu}\left(  k_{2},k_{1}\right)  }{d\left(  k_{2}%
,k_{1}\right)  -x_{2}^{2}m^{2}}+\frac{F_{1}^{\mu}\left(  k_{2},k_{1}\right)
}{D\left(  k_{2},k_{1}\right)  -x_{2}m^{2}}\right)  \frac{1}{\hat{k}_{\bot}%
-m}-K_{1}^{\mu}(k_{2},k_{1})\right\}  \\
&  -t^{\mathcal{G}_{2}}t^{\mathcal{G}_{1}}t^{G_{2}}K_{2}^{\mu}(k_{2}%
,k_{1}^{\prime\prime})+t^{\mathcal{G}_{1}}t^{\mathcal{G}_{2}}t^{G_{2}}%
K_{2}^{\mu}(k_{2},k_{1}^{\prime})
\end{align*}
\vspace{-0.9cm}
\begin{align}
&  -\frac{F_{3}^{\mu}\left(  k_{2},k_{1}\right)  }{d\left(  k_{2}%
,k_{1}\right)  -x_{2}^{2}m^{2}}\frac{t^{G_{2}}t^{\mathcal{G}_{2}%
}t^{\mathcal{G}_{1}}}{\hat{q}_{2\bot}-\hat{r}_{1\bot}-m}+\frac{F_{3}^{\mu
}\left(  k_{2},k_{1}\right)  }{d\left(  k_{2},k_{1}\right)  -x_{2}^{2}m^{2}%
}\frac{t^{G_{2}}t^{\mathcal{G}_{1}}t^{\mathcal{G}_{2}}}{\hat{q}_{2\bot}%
-\hat{r}_{2\bot}-m}\nonumber\\
&  -\frac{F_{1}^{\mu}\left(  k_{2},k_{1}^{\prime\prime}\right)  }{D\left(
k_{2},k_{1}^{\prime\prime}\right)  -x_{2}m^{2}}\frac{t^{\mathcal{G}_{2}%
}t^{G_{2}}t^{\mathcal{G}_{1}}}{\hat{q}_{2\bot}-\hat{r}_{1\bot}-m}+\frac
{F_{1}^{\mu}\left(  k_{2},k_{1}^{\prime}\right)  }{D\left(  k_{2}%
,k_{1}^{\prime}\right)  -x_{2}m^{2}}\frac{t^{\mathcal{G}_{1}}t^{G_{2}%
}t^{\mathcal{G}_{2}}}{\hat{q}_{2\bot}-\hat{r}_{2\bot}-m}\nonumber\\
&  -\left[  t^{G_{2}}t^{\mathcal{G}_{2}}\right]  t^{\mathcal{G}_{1}}\left(
\frac{F_{1}^{\mu}\left(  k_{2}^{\prime},k_{1}\right)  }{D\left(  k_{2}%
^{\prime},k_{1}\right)  -x_{2}m^{2}}\frac{1}{\hat{q}_{2\bot}-\hat{r}_{1\bot
}-m}-K_{1}^{\mu}\left(  k_{2}^{\prime},k_{1}\right)  \right)  \nonumber
\end{align}
\vspace{-0.9cm}
\begin{align*}
&  +\left.  \left[  t^{G_{2}}t^{\mathcal{G}_{1}}\right]  t^{\mathcal{G}_{2}%
}\left(  \frac{F_{1}^{\mu}\left(  k_{2}^{\prime\prime},k_{1}\right)
}{D\left(  k_{2}^{\prime\prime},k_{1}\right)  -x_{2}m^{2}}\frac{1}{\hat
{q}_{2\bot}-\hat{r}_{2\bot}-m}-K_{1}^{\mu}\left(  k_{2}^{\prime\prime}%
,k_{1}\right)  \right)  \right)  \\
&  (\hat{q}_{2\bot}-m)\,.
\end{align*}
The sum of the two last expressions precisely gives the r.h.s. of
(\ref{bootstrapQG}).

\subsection{Antiquark--gluon jet production in central region}

We denote the momenta of the emitted antiquark and gluon as $k_{1}$ and
$k_{2}$ respectively.\smallskip\ The bootstrap condition has the form:%
\begin{align}
&  \langle\mathcal{G}_{1}\mathcal{Q}_{2}|\,\widehat{\{\bar{Q}_{1}G_{2}%
\}}\mathcal{\,}|\mathcal{G}_{\omega}(q_{2\bot})\rangle\,g\,q_{2\bot}%
^{2}+\langle\mathcal{G}_{1}\mathcal{Q}_{2}|\{Q_{1}\bar{G}_{2}\}\mathcal{G}%
_{2}\rangle\nonumber\\
&  =\langle\mathcal{G}_{1}\mathcal{Q}_{2}|\mathcal{Q}_{\omega}(q_{1\bot
})\rangle\mathcal{\,}g\,\gamma_{\mathcal{Q}_{2}\mathcal{G}_{2}}^{\{\bar{Q}%
_{1}G_{2}\}},\label{bootstrap_barQG}%
\end{align}
where $\widehat{\{\bar{Q}_{1}G_{2}\}}\mathcal{\,\,}$ is the operator of
antiquark--gluon production with the matrix element%
\begin{align}
&  \langle\mathcal{G}_{1}\mathcal{Q}_{2}|\,\widehat{\{\bar{Q}_{1}G_{2}%
\}}\mathcal{\,\,}|\mathcal{G}_{1}^{\prime}\mathcal{G}_{2}^{\prime}%
\rangle=\delta(q_{1\bot}+k_{\bot}-q_{2\bot})\nonumber\\
&  \times\left(  \gamma_{\mathcal{Q}_{2}\mathcal{G}_{2}^{\prime}}^{\{\bar
{Q}_{1}G_{2}\}}\delta_{\mathcal{G}_{1}\mathcal{G}_{1}^{\prime}}\delta
(r_{1\bot}-r_{1\bot}^{\prime})r_{1\bot}^{2}+\gamma_{\mathcal{G}_{1}%
\mathcal{G}_{1}^{\prime}}^{G_{2}}\gamma_{\mathcal{Q}_{2}\mathcal{G}%
_{2}^{\prime}}^{\bar{Q}_{1}}\delta\left(  r_{2\bot}+k_{1\bot}-r_{2\bot
}^{\prime}\right)  \right)
\end{align}
and%
\begin{align*}
&  \langle\mathcal{G}_{1}\mathcal{Q}_{2}|\{Q_{1}\bar{G}_{2}\}\mathcal{G}%
_{2}\rangle=\delta(q_{1\bot}+k_{\bot}-q_{2\bot})\\
&  \times\left[  \frac{1}{2k^{-}}\left(  \sum_{G}\Gamma_{\{\bar{Q}_{1}%
G_{2}\}G}^{\mathcal{Q}_{2}}\,\gamma_{\mathcal{G}_{1}\mathcal{G}_{2}}^{G}%
-\sum_{Q}\gamma_{Q}^{\mathcal{Q}_{2}\mathcal{G}_{2}}\underline{\Gamma
}_{Q\{Q_{1}\bar{G}_{2}\}}^{\mathcal{G}_{1}}\,\right)  \right.
\end{align*}
\vspace{-0.9cm}
\begin{align}
&  +\frac{1}{2k_{1}^{-}}\left(  \sum_{G}\Gamma_{\bar{Q}_{1}G}^{\mathcal{Q}%
_{2}}\,\gamma_{\mathcal{G}_{1}\mathcal{G}_{2}}^{\{GG_{2}\}}-\sum_{Q}%
\gamma_{\{Q\bar{G}_{2}\}}^{\mathcal{Q}_{2}\mathcal{G}_{2}}\underline{\Gamma
}_{Q\,Q_{1}}^{\mathcal{G}_{1}}\right)  \nonumber\\
&  +\left.  \frac{1}{2k_{2}^{-}}\left(  \sum_{Q}\Gamma_{G_{2}Q}^{\mathcal{Q}%
_{2}}\,\gamma_{\mathcal{G}_{1}\mathcal{G}_{2}}^{\{\bar{Q}_{1}Q\}}-\sum
_{G}\underline{\Gamma}_{G\,\bar{G}_{2}}^{\mathcal{G}_{1}}\,\gamma_{\{Q_{1}%
G\}}^{\mathcal{Q}_{2}\mathcal{G}_{2}}\right)  \right]  .
\end{align}

The contribution of $\gamma_{\mathcal{G}_{1}\mathcal{G}_{1}^{\prime}}^{G_{2}%
}\gamma_{\mathcal{Q}_{2}\mathcal{G}_{2}^{\prime}}^{\bar{Q}_{1}}$ into
$\langle\mathcal{G}_{1}\mathcal{Q}_{2}|\,\widehat{\{\bar{Q}_{1}G_{2}%
\}}\mathcal{\,}|\mathcal{G}_{\omega}(q_{2\bot})\rangle\,g\,q_{2\bot}^{2}$
yields%
\begin{align}
&  -\delta(q_{1\bot}+k_{\bot}-q_{2\bot})\,g^{3}\left[  \left[  t^{\mathcal{G}%
_{1}}t^{G_{2}}\right]  t^{\mathcal{G}_{2}}\right]  \,\frac{q_{2\bot}^{2}%
}{2k_{1}^{-}}\nonumber\\
&  \times\left(  \frac{{}}{{}}\tilde{K}_{3}^{\mu}\left(  k_{2}^{\prime\prime
},k_{1}^{\prime\prime}\right)  +K_{3}^{\mu}(k_{2}^{\prime\prime},k_{1}%
^{\prime\prime})-\tilde{K}_{3}^{\mu}\left(  k_{2},k_{1}^{\prime\prime}\right)
-K_{3}^{\mu}(k_{2},k_{1}^{\prime\prime})\right)  \gamma^{\mu}\upsilon_{\bar
{Q}_{1}}.
\end{align}
We can present the result for $\langle\mathcal{G}_{1}\mathcal{Q}_{2}%
|\{Q_{1}\bar{G}_{2}\}\mathcal{G}_{2}\rangle$ in the following form:%
\begin{align*}
&  \langle\mathcal{G}_{1}\mathcal{Q}_{2}|\{Q_{1}\bar{G}_{2}\}\mathcal{G}%
_{2}\rangle\\
&  =\delta(q_{1\bot}+k_{\bot}-q_{2\bot})\frac{g^{3}}{k^{-}}\left[
-\frac{\left[  t^{\mathcal{G}_{2}}t^{\mathcal{G}_{1}}\right]  t^{G_{2}%
}q_{2\bot}^{2}}{(q_{2}-r_{1})_{\bot}^{2}}\frac{(q_{2}-r_{1})_{\bot}^{\mu}%
F_{2}^{\mu}(k_{2},k_{1})}{d\left(  k_{2},k_{1}\right)  -x_{2}^{2}m^{2}%
}\right.  \\
&  -t^{\mathcal{G}_{1}}t^{\mathcal{G}_{2}}t^{G_{2}}\frac{q_{2\bot}^{\mu}%
F_{2}^{\mu}(k_{2},k_{1})}{d\left(  k_{2},k_{1}\right)  -x_{2}^{2}m^{2}}%
+\frac{t^{G_{2}}\left[  t^{\mathcal{G}_{1}}t^{\mathcal{G}_{2}}\right]  \hat
{e}_{2}\,\,q_{2\bot}^{2}}{(q_{2}-r_{1})_{\bot}^{2}}\left(  1+\frac
{F_{5}\left(  k_{1},k_{2}^{\prime}\right)  }{D\left(  k_{2}^{\prime}%
,k_{1}\right)  -m^{2}}\right)
\end{align*}
\vspace{-0.9cm}
\begin{align*}
&  +t^{\mathcal{G}_{1}}\left[  t^{G_{2}}t^{\mathcal{G}_{2}}\right]
\frac{\gamma^{\mu}}{2x_{1}}\left(  -\frac{2V_{2}^{\mu}\left(  k_{2}%
,q_{2}-k_{2}\right)  }{D\left(  k_{2},q_{2}-k_{2}\right)  }+\tilde{K}_{3}%
^{\mu}\left(  k_{2},q_{2}-k_{2}\right)  \,q_{2\bot}^{2}\right)  \\
&  -t^{\mathcal{G}_{1}}t^{G_{2}}t^{\mathcal{G}_{2}}\hat{e}_{2}\left(
1+\frac{F_{5}\left(  k_{1},q_{2}-k_{1}\right)  }{D\left(  q_{2}-k_{1}%
,k_{1}\right)  -m^{2}}\right)  +\left[  t^{\mathcal{G}_{1}}\left[
t^{\mathcal{G}_{2}}t^{G_{2}}\right]  \right]  \frac{\gamma^{\mu}q_{2\bot}^{2}%
}{2x_{1}}\tilde{K}_{3}^{\mu}\left(  k_{2},k_{1}^{\prime\prime}\right)  \,
\end{align*}
\vspace{-0.9cm}
\begin{align}
&  -\left[  \left[  t^{G_{2}}t^{\mathcal{G}_{1}}\right]  t^{\mathcal{G}_{2}%
}\right]  \frac{\gamma^{\mu}}{2x_{1}}\left(  \frac{{}}{{}}\tilde{K}_{3}^{\mu
}\left(  k_{2}^{\prime\prime},k_{1}^{\prime\prime}\right)  \,+K_{3}^{\mu
}(k_{2}^{\prime\prime},k_{1}^{\prime\prime})-K_{3}^{\mu}(k_{2},k_{1}%
^{\prime\prime})\right)  q_{2\bot}^{2}\nonumber\\
&  +\left.  \left[  t^{G_{2}}\left[  t^{\mathcal{G}_{1}}t^{\mathcal{G}_{2}%
}\right]  \right]  \frac{\gamma^{\mu}}{x_{1}}\frac{V_{2}^{\mu}\left(
k_{2},k_{1}^{\prime\prime}\right)  }{D\left(  k_{2},k_{1}^{\prime\prime
}\right)  }\,\frac{q_{2\bot}^{2}}{(q_{2}-r_{1})_{\bot}^{2}}\right]
\upsilon_{\bar{Q}_{1}}.
\end{align}
The contribution of $\langle\mathcal{G}_{1}\mathcal{Q}_{2}|\,\widehat
{\{\bar{Q}_{1}G_{2}\}}\mathcal{\,}|\mathcal{G}_{\omega}(q_{2\bot}%
)\rangle\,g\,q_{2\bot}^{2}$ into the bootstrap relation reads as follows:%
\begin{align*}
&  \langle\mathcal{G}_{1}\mathcal{Q}_{2}|\,\widehat{\{\bar{Q}_{1}G_{2}%
\}}\mathcal{\,}|\mathcal{G}_{\omega}(q_{2\bot})\rangle\,g\,q_{2\bot}^{2}\\
&  =\delta(q_{1\bot}+k_{\bot}-q_{2\bot})\frac{g^{3}}{k^{-}}\left[
\frac{\left[  t^{\mathcal{G}_{2}}t^{\mathcal{G}_{1}}\right]  t^{G_{2}%
}\,q_{2\bot}^{2}}{(q_{2}-r_{1})_{\bot}^{2}}\frac{(q_{2}-r_{1})_{\bot}^{\mu
}F_{2}^{\mu}(k_{2},k_{1})}{d\left(  k_{2},k_{1}\right)  -x_{2}^{2}m^{2}%
}\right.
\end{align*}
\vspace{-0.9cm}
\begin{align}
&  -t^{G_{2}}\left[  t^{\mathcal{G}_{1}}t^{\mathcal{G}_{2}}\right]  \hat
{e}_{2}\left(  1+\frac{F_{5}\left(  k_{1},k_{2}^{\prime}\right)  }{D\left(
k_{2}^{\prime},k_{1}\right)  -m^{2}}\right)  \frac{q_{2\bot}^{2}}{(q_{2}%
-r_{1})_{\bot}^{2}}\nonumber\\
&  -\frac{\left[  t^{G_{2}}\left[  t^{\mathcal{G}_{1}}t^{\mathcal{G}_{2}%
}\right]  \right]  \,q_{2\bot}^{2}\gamma^{\mu}}{x_{1}(q_{2}-r_{1})_{\bot}^{2}%
}\frac{V_{2}^{\mu}\left(  k_{2},k_{1}^{\prime\prime}\right)  }{D\left(
k_{2},k_{1}^{\prime\prime}\right)  }\,-\frac{\left[  t^{\mathcal{G}_{1}%
}\left[  t^{\mathcal{G}_{2}}t^{G_{2}}\right]  \right]  \,\,\gamma^{\mu
}\,q_{2\bot}^{2}}{2x_{1}}\tilde{K}_{3}^{\mu}\left(  k_{2},k_{1}^{\prime\prime
}\right)  \nonumber\\
&  -\left.  \frac{\left[  \left[  t^{\mathcal{G}_{1}}t^{G_{2}}\right]
t^{\mathcal{G}_{2}}\right]  q_{2\bot}^{2}\,\gamma^{\mu}}{2x_{1}}\left(
\frac{{}}{{}}\tilde{K}_{3}^{\mu}\left(  k_{2}^{\prime\prime},k_{1}%
^{\prime\prime}\right)  \,+K_{3}^{\mu}(k_{2}^{\prime\prime},k_{1}%
^{\prime\prime})-K_{3}^{\mu}(k_{2},k_{1}^{\prime\prime})\right)  \right]
\upsilon_{\bar{Q}_{1}}.
\end{align}
The sum of the two last expressions gives the r.h.s. of (\ref{bootstrap_barQG}%
). It concludes the proof of the bootstrap conditions.

\section{Summary}

The further development of the quark Reggeization theory in QCD demands a
proof of the quark Reggeization hypothesis in the NLO. Besides the MRK the NLA
also includes another, quasi--multi--Regge kinematics, in which one of the
produced particles is replaced by a jet containing two particles with similar
rapidities. In this paper we have proved the quark Reggeization in the QMRK
\ by means of the method analogous to the proof performed in the LO. It is
based on explicit verification of the so--called bootstrap conditions --- the
constraints on the effective Reggeon vertices. These conditions are imposed by
the bootstrap relations which are required by the compatibility of the
$s$--channel unitarity with the QMRK form of amplitude (\ref{inelastic quark}%
). We formulate these conditions in the operator
formalism in the transverse momentum, colour and spin
space. This formalism was firstly introduced in
\cite{FFKP}, then extended to inelastic amplitudes in
\cite{Bogdan:2006af} and adopted here for the QMRK. The
direct insertion of the effective vertices into the
bootstrap conditions leads to extremely cumbersome and
tedious calculations, which are almost completely
cancellations. We make these cancellations transparent
for verification introducing nontrivial parametrization
of the Reggeon vertices through the set of functions
$F_{i}$, $K_{i}$,.$V_{i}$ (\ref{func1}-\ref{func2}).

\begin{ack}
We would like to thank V.S. Fadin for draw our attention
to this work and for helpful comments and discussions.
\end{ack}


\begin{thebibliography}{99}                                                                                               %


\bibitem {FS}
V.S. Fadin and V.E. Sherman, Pis'ma Zh. Eksp. Teor. Fiz. \textbf{23} (1976)
599 [Sov. Phys. JETP Lett. \textbf{23} (1976) 548]; Zh. Eksp. Teor. Fiz.
\textbf{72} (1977) 1640 [Sov. Phys. JETP \textbf{45} (1977) 861].

\bibitem {BD-DFG}
A.V. Bogdan, V. Del Duca, V.S. Fadin, and E.W.N. Glover, JHEP \textbf{203}
(2002) 32, hep-ph/0201240.

\bibitem {Bogdan_F_1}
A.V. Bogdan and V.S. Fadin, Yad. Phys. \textbf{9} 1659 (2005).

\bibitem {FF01}
V.S. Fadin and R. Fiore, Phys. Rev. D \textbf{64} (2001) 114012, hep-ph/0107010.

\bibitem {KLPV}
M.I. Kotsky, L.N. Lipatov, A. Principe, and M.I. Vyazovsky, Nucl. Phys. B
\textbf{648} (2003) 277, hep-ph/0207169.

\bibitem {Bogdan:2006af}A.~V.~Bogdan and V.~S.~Fadin,
Nucl.\ Phys.\ B \textbf{740}, 36 (2006) [arXiv:hep-ph/0601117].




\bibitem {Lipatov:2000se}L.~N.~Lipatov and M.~I.~Vyazovsky,
Nucl.\ Phys.\ B \textbf{597}, 399 (2001) [arXiv:hep-ph/0009340].


\bibitem {FFKP}
V.~S.~Fadin, R.~Fiore, M.~I.~Kotsky, and A.~Papa,
Phys. Lett. B \textbf{495} (2000) 329, hep-ph/0008057.



\bibitem {Fadin:2003av}V.~S.~Fadin,
Phys.\ Atom.\ Nucl.\ \textbf{66}, 2017 (2003).


\bibitem {BFKL}V.S. Fadin, E.A. Kuraev, and L.N. Lipatov, Phys. Lett. B
\textbf{60} (1975) 50; E.A. Kuraev, L.N. Lipatov, and V.S. Fadin, Zh. Eksp.
Teor. Fiz. \textbf{71} (1976) 840 [Sov. Phys. JETP \textbf{44} (1976) 443];
Zh. Eksp. Teor. Fiz. \textbf{72} (1977) 377 [Sov. Phys. JETP \textbf{45}
(1977) 199].



\bibitem {Fadin:2003xs}V.~S.~Fadin, M.~G.~Kozlov and A.~V.~Reznichenko,
Phys.\ Atom.\ Nucl.\ \textbf{67}, 359 (2004) [Yad.\ Fiz.\ \textbf{67}, 377
(2004)] [arXiv:hep-ph/0302224].

\end{thebibliography}
\end{document}